\documentclass[a4paper,fleqn]{cas-dc}

\usepackage[numbers,sort&compress]{natbib}
\usepackage{subcaption}
\usepackage{lineno}

\usepackage{xcolor}
\usepackage{hyperref}
\hypersetup{citecolor=teal,linkcolor=purple}
\usepackage[normalem]{ulem}

\definecolor{orange}{RGB}{255,127,0}

\begin{document}
\let\WriteBookmarks\relax
\def\floatpagepagefraction{1}
\def\textpagefraction{.001}

\shorttitle{Primordial Asymmetries, Primordial Equation of State \& Primordial Black Holes}    

\shortauthors{Gonin et al.}  

\title[mode=title]{Primordial Asymmetries, Primordial Equation of State \& Primordial Black Holes}  



%

\author[1,2]{Ma\"el Gonin}[orcid=0009-0008-2685-3497]

\cormark[1]


\ead{mael.gonin@dzastro.de}


\credit{Conceptualization, Methodology, Software, Formal analysis, Investigation, Data Curation, Writing - Original Draft, Writing - Review \& Editing, Visualization, Project administration}

\affiliation[1]{organization={Deutsches Zentrum f\"ur Astrophysik (DZA)},
            addressline={Postplatz 1}, 
            city={G\"orlitz},
            postcode={02826 }, 
            country={Germany}}

\affiliation[2]{organization={Institut für Kern- und Teilchenphysik (IKTP), TU Dresden},
            addressline={Zellescher Weg 19}, 
            city={Dresden},
            postcode={01062}, 
            country={Germany}}

\author[3]{Julien Froustey}[orcid=0000-0002-6466-8232]
\ead{julien.froustey@ific.uv.es}

\affiliation[3]{organization={Institut de Física Corpuscular (IFIC), CSIC-Universitat de València},
            addressline={Parc Científic UV, C/ Catedrático José Beltrán 2}, 
            city={Paterna (València)},
          citysep={}, 
            postcode={46980}, 
            country={Spain}}

\credit{Writing - Original Draft, Writing - Review \& Editing, Methodology, Software, Visualization, Funding acquisition}

\author[4,5,6]{Albert Escrivà}[orcid=0000-0001-5483-8034]
\ead{alberto.escriva@apctp.org}
\affiliation[4]{
    organization={Asia Pacific Center for Theoretical Physics},
    city={Pohang},
    postcode={37673},
    country={Republic of Korea}
}
\affiliation[5]{
    organization={Department of Physics, Pohang University of Science and Technology},
    city={Pohang},
    postcode={37673},
    country={Republic of Korea}
}
\affiliation[6]{
    organization={Department of Physics, Nagoya University},
    addressline={Furo-cho, Chikusa-ku},
    city={Nagoya},
    postcode={464-8602},
    country={Japan}
}

\credit{Writing - Original Draft, Writing - Review \& Editing, Methodology, Software, Resources, Visualization, Funding acquisition}

\author[7]{Alberto Magaraggia}[orcid=0009-0009-6060-9540]
\ead{axm8568@miami.edu}
\affiliation[7]{organization={Department of Physics, University of Miami},
            city={Coral Gables},
            postcode={FL 33124}, 
            state={Florida},
            country={USA}}

\credit{Writing - Original Draft, Software, Visualization}

\author[8,9]{Florian K\"uhnel}[orcid=0000-0002-1528-1920]
\ead{fkuehnel@mpp.mpg.de}

\affiliation[8]{
            organization={Fakult{\"a}t f{\"u}r Physik, Technische Universit{\"a}t Dortmund},
            addressline={August-Schmidt-Str. 4}, 
            city={Dortmund},
            postcode={44227}, 
            country={Germany}}
\affiliation[9]{
            organization={Arnold Sommerfeld Center,
        	Ludwig-Maximilians-Universit{\"a}t},
            addressline={Theresienstr.~37}, 
            city={M{\"u}nchen},
            postcode={80333}, 
            country={Germany}}

\credit{Validation, Writing - Original Draft, Writing - Review \& Editing, Funding acquisition}

\author[1,2]{G\"unther Hasinger}[orcid=0000-0002-0797-0646]
\ead{guenther.hasinger@dzastro.de}

\credit{Supervision, Validation, Funding acquisition}

\cortext[1]{Corresponding author}



\begin{abstract}
We study the thermal history of the primordial Universe in  the presence of non-zero lepton and baryon asymmetries. Considering different scenarios, we determine the equation of state (EoS) of the Universe from $T=10 \rm ~GeV$ down to $T = 1 \rm ~keV$, spanning the QCD transition, hadron gas phase and neutrino decoupling epochs. Using a combination of numerical codes, we track the cosmic trajectories of chemical potentials associated with the baryonic, leptonic and electric charges, and follow the evolution of lepton asymmetries including through the era where neutrino oscillations take place. Combining peak theory with numerical-relativity simulations of the collapse threshold, we show the EoS-induced modifications to the primordial black hole (PBH) mass spectrum. We determine the associated Gravitational Wave (GW) signal, showing how lepton asymmetries and a particular spectral index of curvature perturbations can be hinted at by current ground interferometer-based GW observations. Finally, we discuss constraints and positive evidence for PBHs.
\end{abstract}





\maketitle

\section{Introduction}\label{sec:intro}
The first gravitational wave detection by LIGO~\cite{LIGOScientific:2016aoc} opened a new 
era for astronomy and cosmology. This event, together with subsequent detections from the 
four LIGO--Virgo-- KAGRA (LVK) observing runs \cite{LIGOScientific:2026wfs, LIGOScientific:2026pop}, 
revitalised primordial black holes (PBHs) as dark matter 
(DM) candidates \cite{Bird:2016dcv,Sasaki_2018,Clesse:2020ghq,Escriva:2022bwe,Franciolini:2022QCD,DeLuca:2025fln,Magaraggia_2026}. 
These black holes are thought to form during the radiation era 
($z > z_{\rm eq} \sim 3400$) from the gravitational collapse of large-amplitude primordial curvature perturbations \cite{1974MNRAS.168..399C,1975ApJ...201....1C} (see \cite{Escriva:2022duf} for a recent review and list of other mechanisms). 
PBHs play a dual role in cosmology: they could provide explanations for as yet unexplained observations \cite{Carr:2019kxo} and probe fundamental physics 
from the pre-recombination era \cite{Byrnes:2018clq,Bodeker:2020stj,Hashino:2022tcs,Escriva:2022bwe,Musco:2023dak,Gonin:2026xhe}, beyond the reach of current experimental capabilities.

The pre-recombination Universe is an era of many speculations; its evolution is directly linked to our understanding of fundamental physics \cite{Kolb:1990vq}. Although the hot Big Bang model starts at temperatures much higher than the ones that can be reached in ground-based experiments, the Standard Model of particle physics provides the ingredients of our description of the radiation era. In addition, in order to solve some of the problems of the standard Big-Bang model, the radiation era is conventionally preceded by an inflation era. This period of accelerated expansion leaves the Universe almost perfectly flat and provides a source of density perturbations~\cite{Guth:1980zm}. The size of the density fluctuations relative to the Hubble radius is key to 
understanding PBH formation: as the comoving Hubble radius grows, a sufficiently large fluctuation on the scale re-entering the horizon can collapse and form a PBH~\cite{1974MNRAS.168..399C,
1975ApJ...201....1C} (see Fig.~6 of Ref.~\cite{Gonin:2025uvc} for a schematic view). 
The exact inflationary model and the resulting 
fluctuation spectrum (with large overdensities on small scales, needed for PBH formation) are currently subject to theoretical uncertainties.

After the reheating phase (except in very low reheating temperature models, see~\cite{Barbieri:2025moq} for recent constraints), at the electroweak scale $T \sim m_H$, when particles acquire their masses, the 
temperature is close enough to that reached in particle colliders that we can learn about 
the nature of the phase transition (PT)~\cite{Pandav:2025sqo}. Although the conditions in 
particle colliders are not quite the same --- the early Universe evolves with a very small 
net baryon number $n_B = bs$, with $s$ the entropy density and $b = 8.6 \times 10^{-11}$ the baryon asymmetry 
parameter inferred from~\cite{Planck:2018vyg}, whereas colliders reach high temperatures 
with a very large net baryon number --- the comparison remains instructive.

The electroweak PT is thought to be the era of baryogenesis and leptogenesis, when the 
primordial lepton and baryon asymmetries of the Universe (LAU and BAU) are 
set~\cite{Bodeker:2020ghk,vandeVis:2025efm}. However, the measured Higgs boson mass 
$m_H = 125~{\rm GeV}$ from~\cite{ParticleDataGroup:2022pth} suggests a smooth crossover~\cite{Kajantie:1996mn,Kajantie:1996qd}. 
Without a first-order PT, the Sakharov conditions \cite{Sakharov:1967dj} necessary for baryogenesis and 
leptogenesis appear hard to fulfill. Nevertheless, 
the motivation to find a first-order PT remains, as one needs a mechanism to explain the 
observed baryon asymmetry $b = 8.6 \times 10^{-11}$~\cite{Khlopov:2021xnw}. Moreover, 
the latest Big Bang Nucleosynthesis (BBN) and Cosmic Microwave Background (CMB) studies suggest that the lepton asymmetry parameter $\ell = \ell_e + 
\ell_\mu + \ell_\tau$, defined similarly to $b$ [see Eq.~\eqref{eq:conservation_equations:a} below] could be as large as $\mathcal{O}(10^{-2})$~\cite{Oldengott:2017tzj,
Kawasaki:2022hvx,Escudero:2022okz,Froustey:2024mgf,Lattanzi:2024hnq,Li:2024gzf,Domcke:2025lzg,
Domcke:2025jiy}. Its exact value, and especially the individual flavour asymmetry values, remain elusive due to the collective neutrino oscillations occurring before BBN.

Another relevant cosmic PT is the QCD transition, when the phase of strongly interacting 
matter shifts from the quark-gluon plasma to a hadron gas. Lattice QCD simulations, working at vanishing 
chemical potential, suggest a smooth crossover \cite{Aoki:2006we,Borsanyi:2016ksw}. However, it was shown by Schwarz \& 
Stuke~\cite{Schwarz:2009ii} that in the presence of primordial asymmetries $b$ and 
$\ell_\alpha$ (where $\alpha = e, \mu, \tau$), the particle distributions acquire chemical 
potentials $\{\mu_B,\,\mu_Q,\,\mu_{\nu_e},\,\mu_{\nu_\mu},\,\mu_{\nu_\tau}\}$, which 
modify the thermodynamic picture and can, in extreme cases, trigger a first-order QCD 
transition with implications for the stochastic gravitational wave background (SGWB) \cite{Gao:2021nwz,Gao:2023djs,Gao:2024fhm}.

The radiation era is often described through the evolution of the equation of state (EoS) 
parameter $w = P/\varepsilon$ and the squared sound speed $c_s^2 = \partial 
P/\partial\varepsilon$, where $P$ is the total pressure and $\varepsilon$ the total 
energy density. Starting as an extremely hot phase, the Universe expands while cooling 
down; adiabatic expansion relates the entropy density and scale factor as $s \propto a^{-3}$. Prior to the electroweak PT, all particles were relativistic, giving 
$w = 1/3$; as PTs occur, or when species transition to non-relativistic regimes, $w$ 
departs from the radiation value of $1/3$, with $w < 1/3$ associated with a 
\textit{softening} of the Universe. The sound speed $c_s$ follows the same behaviour, 
i.e.\ its value decreases during cosmic PTs~\cite{Byrnes:2018clq,Escriva:2022bwe,Musco:2023dak}. A particle species $i$ becomes 
non-relativistic when $T \sim m_i$; this corresponds to the end of the production of the 
associated antiparticles, and at $T \sim m_i$ the particle $i$ annihilates with its 
antiparticle partner. If a species carries a non-vanishing asymmetry parameter, such 
events depend on its value.

Beyond the Standard Model, smooth thermal crossovers in additional strongly coupled sectors have also been investigated using holographic equations of state, showing that such transitions can modify PBH formation and generate characteristic features in the resulting mass spectrum \cite{Escriva:2022yaf}.

The soft imprints left in the EoS by PTs directly manifest in the PBH mass distribution: 
when considering a broad fluctuation spectrum, the reduction of the sound speed is 
associated with enhanced PBH formation at a characteristic mass scale \cite{Byrnes:2018clq,Carr:2019kxo,
Bodeker:2020stj,Escriva:2022bwe}. In this study we 
follow the evolution of the asymmetries from the QCD transition to neutrino decoupling and 
evaluate the associated PBH mass spectra for various values of LAU using peak theory \cite{Bardeen:1985tr}.
Current BBN+CMB constraints allow for a range of possible $\ell_\alpha$, leading to different EoS realisations beyond the Standard scenario $\ell_\alpha \simeq 0$. We present the variety of resulting PBH mass spectra and 
discuss how these could impact gravitational wave detections.

One of the main challenges facing the science of PBHs is to accurately model the 
evolution of a population across cosmic time. Given the many theoretical uncertainties on 
the initial conditions --- clustering or not? formation from non-Gaussianities? the extent 
of the PBH mass range? --- and their impact on the constraints, PBHs still await 
confirmation or definitive exclusion. As shown here and in~\cite{Byrnes:2018clq,Carr:2019kxo,
Bodeker:2020stj}, the deep connection between the thermal history and the PBH mass 
spectrum is a remarkable feature of nature. Through the cosmic EoS, the mass distribution 
can hint at the fundamental physics of our Universe.

This paper is organised as follows. In Section~\ref{sec:method_EoS} we describe the method 
used to compute the EoS from 10 GeV to the BBN epoch; in Section~\ref{sec:method_PBH} we outline peak theory and 
gravitational wave production from a PBH population. Results are presented in 
Section~\ref{sec:Results}; we discuss them and place them in the context of current 
constraints on PBHs in Section~\ref{sec:discussion}, before concluding in 
Section~\ref{sec:conclusion}. This paper includes three appendices containing figures that 
support the claims of the main text.

Throughout this paper, we use natural units for which $\hbar = k_\mathrm{B} = c = 1$, and the masses of the particles are set according to \cite{ParticleDataGroup:2022pth}.

\section{Methods: determining the cosmic Equation of State}\label{sec:method_EoS}

In order to study how $w(T)$ evolves, one needs to model the primordial plasma over a wide 
temperature range. We focus our study on the period spanning the QCD transition at $T_c \sim 156~{\rm MeV}$, $\mu^+\mu^-$ annihilations at $T_{\mu^\pm} \sim m_\mu = 106 \, \mathrm{MeV}$, neutrino decoupling at $T_{\rm dec} \sim 1~{\rm MeV}$, and $e^+e^-$ annihilations at  $T_{e^\pm} \sim m_e = 0.511~{\rm MeV}$. These transitions produce the most drastic departures from $w = 1/3$~\cite{Borsanyi:2016ksw,Carr:2019kxo,Gonin:2026xhe} and are also the least subject to theoretical uncertainties.  The primordial 
plasma comprises the different fundamental sectors: QCD particles, leptons, weak 
bosons\footnote{At the temperature range considered, their contribution to the 
thermodynamics may be negligible. Only BSM extensions involving low-mass bosons could be 
relevant; see~\cite{Gonin:2025uvc}.} and photons.

Due to the various physical processes involved at different temperature ranges, the problem 
of resolving the EoS is split into three distinct temperature domains, each relying on 
different codes, assumptions, and approximations. 
From $10~{\rm GeV}$ to $37~{\rm MeV}$ we rely on \texttt{CosmicEoS},\footnote{Not yet 
publicly available; we hope to make it so in the near future.} see 
Sec.~\ref{subsec:EoS_QCD}; from $37~{\rm MeV}$ to $10~{\rm MeV}$ we use 
\texttt{Thermal-FIST}~\cite{Vovchenko:2019pjl}, see Sec.~\ref{subsec:HRG}; finally, we 
make use of \texttt{nudec\_BSM}~\cite{EscuderoAbenza:2020cmq,Escudero:2025kej} and \texttt{NEVO}~\cite{Froustey:2020mcq,Froustey:2021azz,Froustey:2022sla} below 10 MeV to describe the neutrino decoupling epoch, see Sec.~\ref{subsec:nu_osc}.

\subsection{The Cosmic Equation of State across the QCD transition: \texttt{CosmicEoS}}\label{subsec:EoS_QCD}

In this section we present the methods behind the code \texttt{CosmicEoS}.
At $T > 30~{\rm MeV}$, neutrinos are 
still in thermal equilibrium with the rest of the plasma, sharing the same 
temperature, and the QCD sector is best described through lattice QCD simulations. These 
allow exploration of the nuclear matter phase diagram in a cosmological 
context~\cite{Borsanyi:2016ksw} (Supplementary Information of that reference); 
see~\cite{Guenther:2020jwe,Guenther:2022wcr} for a recent overview. Alternatively, 
functional QCD methods can be used~\cite{Gao:2021nwz,Gao:2023djs,Gao:2024fhm,Ferreira:2025zeu}; in the present study we use the microscopic model 
of~\cite{Blaschke:2023pqd} based on the generalised Beth-Uhlenbeck (GBU) approach from $10~{\rm MeV}$ to $1300~{\rm MeV}$. Above $2~{\rm GeV}$ 
we use the extrapolation parameters of~\cite{Bresciani:2025vxw}. For a smooth transition 
between the two datasets we use cubic interpolation. 
We rely on a Taylor expansion to evaluate the QCD sector in the presence of asymmetries, 
following an approach similar to~\cite{Schwarz:2009ii,Wygas:2018otj,
Middeldorf-Wygas:2020glx,Formaggio:2025nde}. The details of our method can be found 
in~\cite{Gonin:2026xhe}. We use the code \texttt{CosmicEoS} to obtain the chemical 
potentials $\{\mu_B,\,\mu_Q,\,\mu_{\nu_e},\,\mu_{\nu_\mu},\,\mu_{\nu_\tau}\}$ by solving 
the following conservation equations:
\begin{subequations}
\label{eq:conservation_equations}
\begin{align}
    \ell_\alpha s &= n_\alpha + n_{\nu_\alpha} = n_{L_\alpha} \, , \label{eq:conservation_equations:a}\\
    bs &= \sum_i B_i n_i = n_B \, , \label{eq:conservation_equations:b}\\
    qs &= \sum_i Q_i n_i = n_Q \, , \label{eq:conservation_equations:c}
\end{align}
\end{subequations}
where $n_B$ is the net baryon number density, $n_Q$ the net electric charge number 
density, $n_{L_\alpha}$ the net lepton number density for flavour $\alpha$, $\ell_\alpha$ 
the lepton flavour asymmetry for $\alpha = e,\,\mu,\,\tau$, and $s$ the total entropy 
density. We consider an electrically neutral Universe, $q=0$ \cite{Caprini:2003gz}. We stress that all number densities are net quantities: $n_\alpha = n_{\alpha^-} 
- n_{\alpha^+}$, and similarly for neutrinos. It is convenient to define the asymmetry 
parameter $\eta_i = n_i/T^3$, where $n_i$ is again a net number density. Chemical 
equilibrium then allows us to write the quark and lepton chemical potentials as:
\begin{subequations}
\label{eq:part_chem_pot}
\begin{align}
    \mu_{\text{up-type}} &= \frac{1}{3}\mu_B + \frac{2}{3} \mu_Q , \label{eq:part_chem_pot:a}\\
    \mu_{\text{down-type}} &= \frac{1}{3}\mu_B - \frac{1}{3} \mu_Q , \label{eq:part_chem_pot:b}\\
    \mu_{\nu_\alpha} &= \mu_{\alpha^{\pm}} + \mu_Q ,                          \label{eq:part_chem_pot:c} 
\end{align}
\end{subequations}
where $\mu_Q$ is the electric charge chemical potential, $\mu_{\nu_\alpha}$ is the 
chemical potential associated with neutrino flavour $\alpha$, $\mu_{\alpha^{\pm}}$ that of 
the charged leptons, and $\mu_{\rm up\text{-}type}$ and $\mu_{\rm down\text{-}type}$ 
those of the up-type quarks $(u, c, t)$ and down-type quarks $(d, s, b)$, respectively.
See~\cite{Formaggio:2025nde} for further details. 
The charm quark flavour was included in the QCD sector through a tree-level correction (see the Supplementary Information of Ref.~\cite{Borsanyi:2016ksw}) to 
the base thermodynamics.\footnote{``Base thermodynamics'' refers to the thermodynamic 
quantities that are not Taylor-expanded, i.e., $s(T,{\vec{\mu}=\vec{0}})$ where 
$\vec{\mu} = \{\mu_B,\mu_Q,\mu_{\nu_e},\mu_{\nu_\mu},\mu_{\nu_\tau}\}$.}
The Taylor expansion coefficients, known as susceptibilities, are taken 
from~\cite{Abuali:2025tbd} for a (2+1) quark flavour configuration and 
from~\cite{Kaczmarek:2025dqt} for the charmed susceptibilities. The procedure to determine 
(2+1+1) susceptibilities is detailed in~\cite{Wygas:2019tsx,Formaggio:2025nde}; 
in~\cite{Gonin:2026xhe} we detail how to extend these over a wide temperature range. We 
use the JEL polynomial approach to obtain the charged lepton distributions with chemical 
potentials~\cite{Johns:1996ht}.

\subsection{The Hadron Gas phase: \texttt{Thermal-Fist}}\label{subsec:HRG}
In the temperature range $10 \leq T \leq 37~{\rm MeV}$, the chemical potentials and 
thermodynamics are determined using the hadron resonance gas model \texttt{Thermal-FIST}~\cite{
Vovchenko:2019pjl}. 

Technically, \texttt{CosmicEoS} is limited to a minimum temperature of 
$30~{\rm MeV}$ by the susceptibility dataset of~\cite{Abuali:2025tbd}, such that we also solve the conservation 
equations~(\ref{eq:conservation_equations:a})--(\ref{eq:conservation_equations:c}) with \texttt{CosmicEoS} down to this limit.

Both \texttt{CosmicEoS} and \texttt{Thermal-FIST} exhibit consistent behaviour at 
$T \sim 30~{\rm MeV}$, but a naive concatenation of the two datasets at $T = 30~{\rm MeV}$ 
produces a discontinuity. We find that the two datasets connect most smoothly at 
$T = 37~{\rm MeV}$. As seen in the figures below, the discrepancy between the two datasets 
is small.

\subsection{Neutrino oscillations and electron-positron annihilations: \texttt{NEVO} and \texttt{nudec\_BSM}}\label{subsec:nu_osc}

Below 10 MeV, the assumption of thermal equilibrium fails for neutrinos as they \emph{decouple} from the electromagnetic plasma. The non-relativistic transition of electrons and positrons, at $T_{e^\pm} \sim 0.511 \, \mathrm{MeV}$, leads to an entropy transfer mostly to photons, resulting in a difference in temperature between the photon and neutrino cosmic backgrounds. In order to compute the EoS of the Universe in this temperature range, we must thus describe the process of neutrino decoupling, which for instance allows one to predict the value of the parameter $N_\mathrm{eff} = 3.044$ in the standard, lepton-symmetric, case~\cite{Akita:2020szl,Froustey:2020mcq,Bennett:2020zkv}. In this standard case with vanishing asymmetries, the non-thermal features of the neutrino distributions remain small, such that an approximate description assuming that neutrinos maintain a thermal distribution (at a temperature different from the photon one) produces sufficiently accurate results. We then use the public code~\texttt{nudec\_BSM}~\cite{EscuderoAbenza:2020cmq,Escudero:2025kej} for this scenario.

In this temperature range, a key feature comes into play with significant consequences on the evolution of asymmetries: neutrino oscillations. Because of flavour mixing, the conservation equation~\eqref{eq:conservation_equations:a} is not valid below 10 MeV. Non-zero LAU actually contribute to the neutrino mixing potential, leading to a non-linear, collective behaviour called \emph{synchronous} neutrino oscillations (see e.g.,~\cite{Samuel:1993uw,Pastor:2001iu,Wong:2002fa,Froustey:2021azz}). In order to model this physics, we use the \texttt{NEVO} code to calculate the evolution of the (anti)neutrino one-body density matrices $\varrho_{\alpha \beta}$ ($\bar{\varrho}_{\alpha \beta}$)\footnote{This object generalises the neutrino distributions $\{f_{\nu_e},f_{\nu_\mu},f_{\nu_\tau}\}$ into a matrix in flavour space $\varrho_{\alpha \beta}$~\cite{Sigl:1993ctk}.} for temperatures smaller than 10 MeV~\cite{Froustey:2020mcq,
Froustey:2021azz,Froustey:2022sla,Froustey:2024mgf}. The results are calculated as a function of the comoving temperature $T_\mathrm{cm} \propto a^{-1}$ (with $a$ the scale factor), which coincides with the common temperature of all species above 10 MeV. We use 40 momentum bins, equally spaced in the range $0 < y = p/T_\mathrm{cm} \leq 24$. The chemical potentials obtained at the end of the \texttt{Thermal-FIST} run allow us to define the initial Fermi-Dirac distributions used in \texttt{NEVO}. We distinguish between normal ordering (NO) and inverted ordering (IO) of the neutrino masses, using the central values of the mixing parameters from~\cite{ParticleDataGroup:2022pth}. No CP-phase is included, as it only leads to an initial dephasing of the fast accelerating, and eventually averaged out, collective oscillations~\cite{Froustey:2021azz}.

Thermodynamic quantities can be calculated from the neutrino distributions in \texttt{NEVO}, for instance
\begin{equation}
    n_{\nu_\alpha}(T_\mathrm{cm}) = \frac{T_\mathrm{cm}^3}{2 \pi^2} \int{\mathrm{d}y \, y^2 \, \left[\varrho_{\alpha \alpha}(T_\mathrm{cm},y) - \bar{\varrho}_{\alpha \alpha}(T_\mathrm{cm},y)\right]} \, .
\end{equation}
The photon and charged lepton distributions are equilibrium ones, at a temperature $T_\gamma \neq T_\mathrm{cm}$ (because of $e^\pm$ annihilations).

\subsection{The evolution of asymmetries and the cosmic trajectories}

It is well-known that neutrino oscillations, by redistributing the different flavour asymmetries, essentially ``wash out'' large, primordial $\ell_\alpha$ to small values compatible with CMB and BBN constraints (see e.g.,~\cite{Dolgov:2002ab,Castorina:2012md,Barenboim:2016shh}). Recent work has actually made this statement much more precise, revealing a more intricate allowed parameter space of primordial asymmetries than previously thought~\cite{Froustey:2024mgf,Domcke:2025lzg,Domcke:2025jiy}. As a consequence, one could consider very large LAU prior to oscillations; however, we are here limited by 
the Taylor expansion invoked in Sec.~\ref{subsec:EoS_QCD}, which remains a viable approximation up to $\mu_B/T \sim 0.1$. 

We choose to consider 4 models of LAU:
\begin{align}
\label{LAU:std}
\begin{aligned}
    b = 8.6\times10^{-11} \, &, \\
    \ell=\ell_e +\ell_\mu+\ell_\tau =-\frac{51}{28}b \, &,
\end{aligned} && &\text{(Standard)} \\[8pt]
\label{LAU:typeI_0.1}
\begin{aligned}
    b=8.6\times10^{-11} \, &, \\
    \ell_e=0 \, , \ \ \ell_\mu=-\ell_\tau=-0.1  \, &,
    \end{aligned} && &\text{(Type I -- 0.1)} \\[8pt]
\label{LAU:typeI_0.08}
\begin{aligned}
    b=8.6\times10^{-11} \, &, \\
    \ell_e=0 \, , \ \ \ell_\mu=-\ell_\tau=-0.08 \, &,
\end{aligned} && &\text{(Type I -- 0.08)} \\[8pt]
\label{LAU:bodeker}
\begin{aligned}
    b=8.6\times10^{-11} \, &, \\
    \ell_e =-0.08 \, , \ \ \ell_\mu=\ell_\tau=0.04 \, &.
\end{aligned} && &\text{(BKOS-like)}
\end{align}

``Standard'' denotes LAU arising from the sphaleron process~\cite{Harvey:1990qw}, with individual flavour asymmetries distributed as $\ell_\alpha=\ell/3$. ``Type I'' follows the prescription of~\cite{Domcke:2025lzg,Ferreira:2025zeu} on the 
preferred direction in the $\ell_\alpha$ parameter space, valid for both NO and IO of neutrino masses. ``BKOS-like''\footnote{BKOS refers to the authors' names: Bödeker, Kühnel, Oldengott and Schwarz.} corresponds to the third model of~\cite{Bodeker:2020stj}, the first existing study on LAU and PBH formation. Using \texttt{NEVO}, we find that this model fulfills the observational 
constraints for a NO of neutrino masses, but is largely excluded in the IO case. We find however that the EoS is extremely similar in both cases (see Sec.~\ref{subsec:EoS_results} and Appendix~\ref{app:details_NO_IO}), such that we do not need to exclude the latter.

We note that setting the same value of $b$ in all four cases is, in principle, inconsistent with the baryon density $\Omega_b h^2$ inferred from the CMB~\cite{Planck:2018vyg}. Indeed, when lepton asymmetries are non-zero, their redistribution by neutrino oscillations is an irreversible process which creates entropy~\cite{Froustey:2024mgf}. As a consequence, the comoving entropy $s a^3$ today is \emph{larger} than at our initial temperature (10 GeV), so the value of $b = n_\mathrm{B}/s$ deduced from~\cite{Planck:2018vyg} is \emph{smaller} than the one we should use in our calculations. Using \texttt{NEVO}, we find that the total entropy increases by 10--20 \% depending on the LAU case. However,  we have verified that these changes of $b$ have negligible effects on the EoS, as expected since those corrections do not change the fact that $b \ll 1$. We therefore keep the same $b$ throughout, also allowing for direct comparison with previous literature.

\begin{figure*}[t]
    \centering
    \includegraphics[trim=1cm 0cm 1cm 0cm, clip, width=\textwidth]{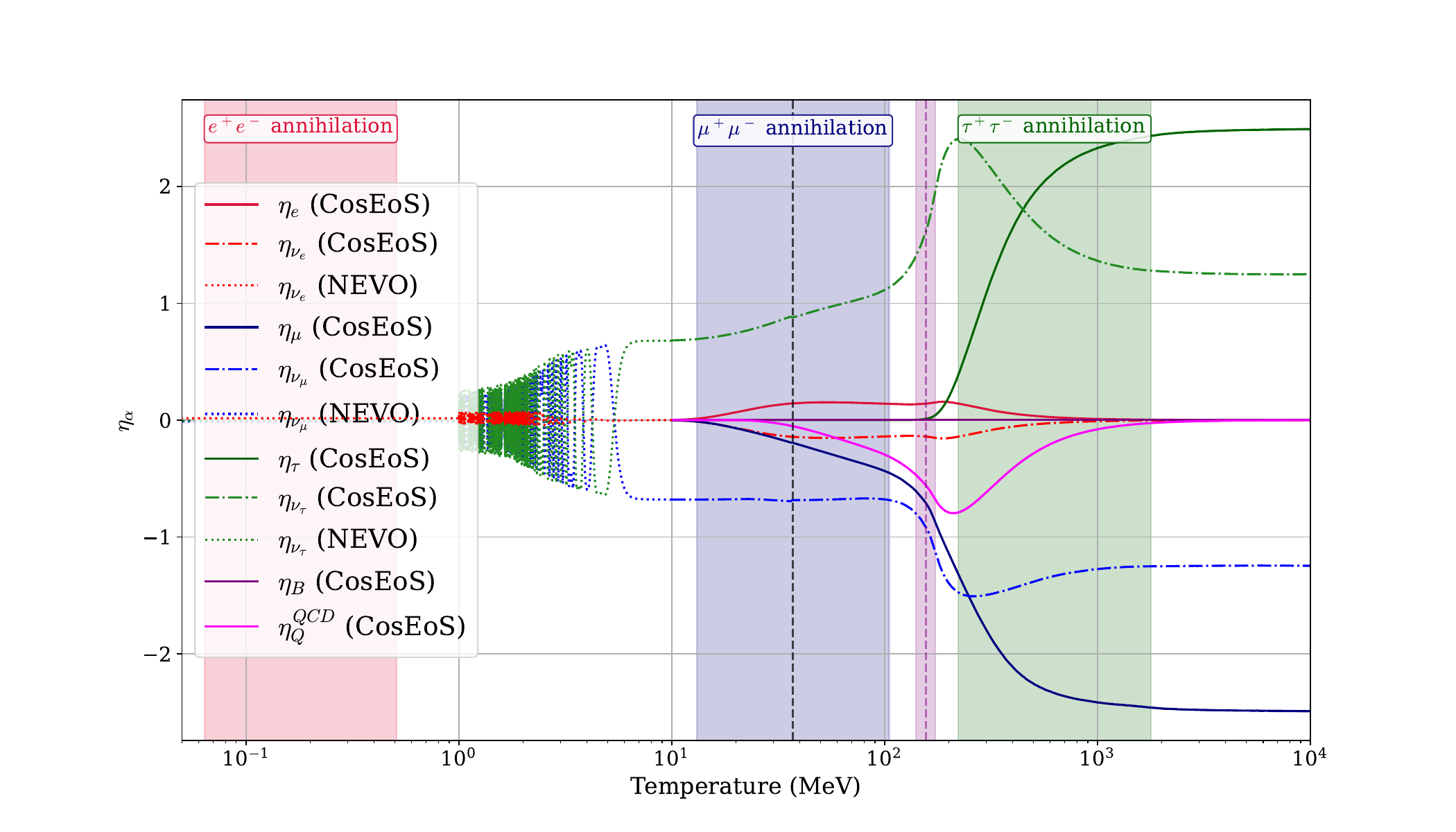} 
    \caption{Evolution of the asymmetry parameter $\eta_i$ across the different regimes for the model 
    ${\rm (Type~I-0.1)}$, see Eq.~\eqref{LAU:typeI_0.1}, with a normal neutrino mass ordering. The shaded regions span from $T = m_\alpha$ to $T = m_\alpha/8$ to highlight the fact that particle/antiparticle annihilation is not 
    instantaneous; at $T = m_\alpha/8$ approximately $90\%$ of the $\alpha$ particles have 
    disappeared. The purple shaded region around the pseudo-critical QCD temperature $T_c = 156.5 \, \mathrm{MeV}$ is purely illustrative. The vertical black dashed line at $T = 37~{\rm MeV}$ marks the 
    junction point between \texttt{CosmicEoS} (shortened as \texttt{CosEoS} in the legend) and \texttt{Thermal-FIST}. The colours are associated to different components: red to the electron sector, blue to the muon sector, green to the tau sector, and purple lines relate to the QCD sector. Solid lines denote electrically charged contributions, while dash-dotted lines show electrically neutral ones. The dotted lines are computed with \texttt{NEVO} (for temperatures below $10 \, \mathrm{MeV}$).} 
    \label{fig:asym_evol}
\end{figure*}

Given the size of the parameter space, we do not attempt an exhaustive survey of the 
possible EoS configurations and their associated PBH mass spectra. Instead, we focus on 
updating the results of~\cite{Bodeker:2020stj} in light of the latest developments in 
neutrino quantum kinetics and PBH mass spectrum evaluations. Additionally, the preferred 
directions of~\cite{Domcke:2025lzg,Ferreira:2025zeu} provide only a partial picture of 
the parameter space, as they are derived by fixing the total LAU to $\ell = 0$. The 
results of~\cite{Froustey:2024mgf,Domcke:2025jiy} have shown that a Universe with $\ell \neq 0$ is indeed physically realisable.  A more comprehensive exploration of the cosmic EoS and PBH formation in the flavour asymmetry directions allowed by BBN+CMB constraints is left for future work.
 
We now discuss the different assumptions between the high-temperature regime $T > 10~{\rm 
MeV}$ and the low-temperature regime $T \lesssim 10~{\rm MeV}$. In \texttt{NEVO}, all 
lepton asymmetries reside in the neutrinos, whereas in \texttt{CosmicEoS} and 
\texttt{Thermal-FIST} the leptonic charge is distributed between the charged leptons and 
their associated neutrinos; see Eq.~(\ref{eq:conservation_equations:a}). In other words, 
\texttt{NEVO} neglects the chemical potentials of the charged leptons. This approach is 
effectively valid for $\ell = \ell_e + \ell_\mu + \ell_\tau \sim 0$ and allows for a relatively smooth connection between 
the two regimes. The reason lies in the electric charge neutrality imposed by 
Eq.~(\ref{eq:conservation_equations:c}): for $b > 0$, the QCD sector can easily carry a 
positive electric charge, while setting $\ell = 0$ forces 
the leptons to compensate their electric charges among themselves when equilibrium is 
fulfilled. Thus, at $T = 10~{\rm MeV}$, when the contributions from the QCD sector, $\tau^\pm$, and $\mu^\pm$ have almost fully vanished, any asymmetry in the $e^\pm$ sector induces 
an electric charge that cannot be cancelled by other species. 
Equation~(\ref{eq:conservation_equations:c}) prevents any departure from charge neutrality, and at $T=10\rm ~MeV$ 
the total asymmetry is effectively set to $\eta_e = \eta_{e^\pm} + \eta_{\nu_e} \sim 
\eta_{\nu_e}$.

We illustrate our multi-code approach by showing the evolution of the parameter $\eta_i = n_i/T^3$ for the model (Type I -- 0.1) in Fig.~\ref{fig:asym_evol}. The behaviour of $\eta_{\alpha^\pm}$ and $\eta_{\nu_\alpha}$ as described in the previous paragraph  appears clearly. The transfer of asymmetry from $\alpha^\pm$ to $\nu_\alpha$ is 
particularly striking for the $\tau^\pm$: as the density of $\tau^\pm$ becomes exponentially suppressed, the associated neutrino $\nu_\tau$ steps up to carry the lepton number. At the same time, due to charge neutrality and $\ell_\mu = -\ell_\tau$, the $\mu^\pm$ must follow the $\tau^\pm$. While the $\tau^\pm$ can no longer carry electric 
charge, the $\mu^\pm$ are still in thermal equilibrium with the remaining species; therefore the 
QCD and $e^\pm$ sectors begin to carry an electric charge as well; see $\eta_{Q}^{\rm 
QCD}$ and $\eta_{e^\pm}$. As $\eta_{e^\pm}$ increases, $\eta_{\nu_e}$ follows with 
opposite sign to keep $\ell_e \sim 0$. Eventually, at $T \sim 200~{\rm MeV} \sim 
m_\tau/8$, the QCD transition forces the start of the disappearance of its associated electric charge; 
then at $T \sim m_\mu = 106~{\rm MeV}$, the $\mu^\pm$ density drops and all electrically charged asymmetries must vanish as well. Below 10 MeV, synchronous oscillations redistribute the flavour asymmetries. At 1 MeV, the \texttt{NEVO} solver switches to an adiabatic approximation which neglects the neutrino self-interaction potential; this effectively averages out the very fast collective oscillations (see details in~\cite{Froustey:2021azz,Froustey:2024mgf}).

Note that in Fig.~\ref{fig:asym_evol} a positive $\eta_{\alpha^\pm}$ actually represents 
a negative electric charge. Comparing the electric charges would require flipping the sign 
of the green, navy blue, and crimson continuous lines. With this in mind, it becomes clear 
that the $\tau^\pm$ transfer their electric charge to the $e^\pm$ and QCD 
sectors, as the $\mu^\pm$ are still ultrarelativistic when the $\tau^\pm$ annihilate and a surplus of positive electric charge exists in the $\mu^\pm$ sector.

The purple continuous line shows $\eta_B$, which is set to very small values by 
Eq.~(\ref{eq:conservation_equations:b}); the absence of any deviation from $\eta_B = 0$ 
demonstrates the self-consistency of \texttt{CosmicEoS}.

\begin{figure}
    \centering
    \includegraphics[trim=1cm 0cm 1cm 1cm, clip, width=\linewidth]{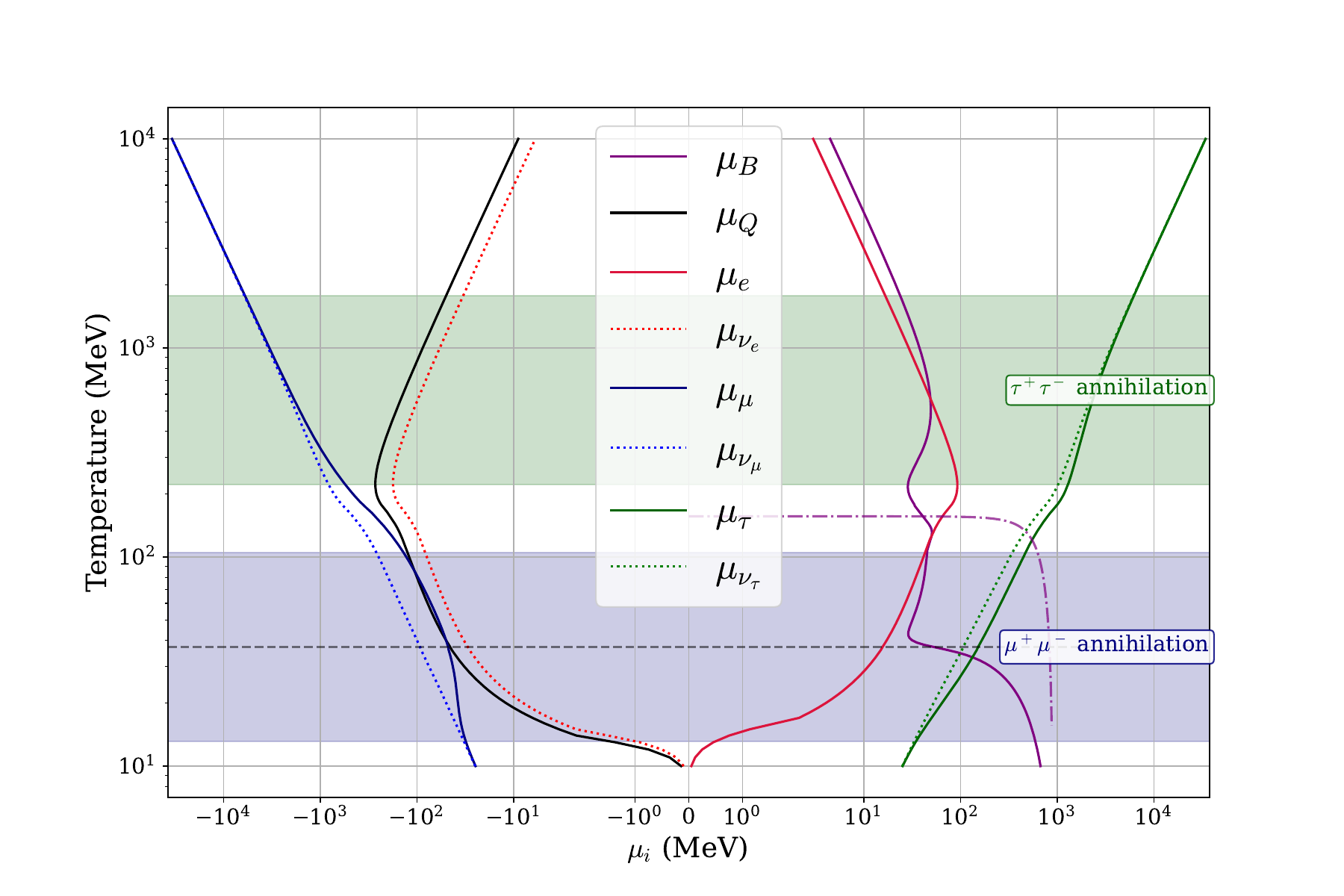}
    \caption{Cosmic trajectories of the different chemical potentials computed with \texttt{CosmicEoS} 
    for the model ${\rm (Type~I-0.1)}$; see Eq.~\eqref{LAU:typeI_0.1}. As in 
    Fig.~\ref{fig:asym_evol}, the shaded regions indicate the annihilation eras, and the 
    horizontal black dashed line at $T = 37~{\rm MeV}$ marks the transition temperature 
    between \texttt{CosmicEoS} and \texttt{Thermal-FIST}. The purple dash-dotted curve shows 
    $T_c(\mu_B)$, the pseudo-critical temperature of the QCD transition, taken 
    from~\cite{Blaschke:2024jqd}.}
    \label{fig:cosmic_traj}
\end{figure}

A complementary viewpoint on the effects of BAU and LAU to the one shown in Fig.~\ref{fig:asym_evol} is the behaviour of the chemical potentials $\{\mu_B, \mu_Q, \mu_{\nu_e}, \mu_{\nu_\mu}, \mu_{\nu_\tau} \}$, which generate the so-called \textit{cosmic 
trajectories}~\cite{Wygas:2018otj,Middeldorf-Wygas:2020glx,Formaggio:2025nde}. 
We show their evolution in Fig.~\ref{fig:cosmic_traj}, obtained 
by solving Eqs.~(\ref{eq:conservation_equations:a})--(\ref{eq:conservation_equations:c}) 
and using Eqs.~(\ref{eq:part_chem_pot:a})--(\ref{eq:part_chem_pot:c}) for the chemical 
potentials of the charged leptons. This evolution is often displayed in the $T~{\rm vs}~
\mu_i$ plane to project the evolution of $\mu_B$ onto the nuclear matter phase diagram. 
The dash-dotted purple line serves as a visual guide, showing the pseudo-critical 
temperature of the QCD transition as a function of $\mu_B$. Schematically, for 
$T > T_c(\mu_B)$ the quarks move freely, while for $T \lesssim T_c(\mu_B)$ they are 
confined.
One can see how the LAU and BAU can shift the trajectory of $\mu_B$ to large values at 
the QCD transition. This is a well-known effect of the LAU and 
BAU~\cite{Schwarz:2009ii,Wygas:2018otj,Formaggio:2025nde,2025arXiv251111995D}; it can change the nature of 
the QCD transition or form a pion condensate \cite{Ferreira:2025zeu}. None of the models discussed here crosses the critical end point (CEP), 
where the QCD transition ceases to be a crossover. 
Refs.~\cite{Gao:2021nwz,Gao:2023djs,Gao:2024fhm,Lu:2023msn} explored these cosmic 
trajectories using a fully functional QCD approach (not relying on Taylor expansion), 
searching for the location of the CEP and for signatures in the SGWB from a first-order 
QCD transition.

Figure~\ref{fig:cosmic_traj} highlights the correlations between the different particle 
species. One can see how $\mu_{\nu_\mu}$ and $\mu_{\nu_\tau}$ evolve together with 
opposite signs, while $\mu_e$ follows an evolution similar to that of $\mu_Q$. One can 
also see how $\mu_B$ exhibits a zig-zag feature at $T \sim 200~{\rm MeV}$ to compensate 
for the residual electric charge from the leptons. The smooth connection between 
\texttt{CosmicEoS} and \texttt{Thermal-FIST} at $T = 37~{\rm MeV}$ validates the 
calculation at low temperatures. At high $T$, we verified that the results tend to the 
ideal gas limit; we choose not to show this in Fig.~\ref{fig:cosmic_traj} to avoid 
overloading the figure.

\section{Methods: Primordial Black Hole mass spectrum \& merger rate}\label{sec:method_PBH}

\subsection{The primordial black hole mass function from peak theory} \label{subsec:method_PBH_mass_function}
PBHs may have formed in the early Universe from the gravitational collapse of large-amplitude primordial curvature perturbations \cite{1974MNRAS.168..399C} (see \cite{Escriva:2022duf} for a recent review and list of other mechanisms). If the amplitude of an overdense region is sufficiently large at horizon re-entry, pressure gradients cannot prevent collapse and a black hole may form. The abundance of PBHs is described by the PBH mass function $f_{\rm PBH}(M)$, which gives the abundance of PBHs as a function of their mass and provides the main connection between early-Universe models and observational constraints.

A crucial ingredient in the calculation of the PBH mass spectrum is the critical threshold for gravitational collapse. This means that the curvature fluctuation amplitude should be above a given threshold for gravity to overcome the pressure gradients and collapse forming a black hole. Since PBH formation is a rare-event process, the predicted abundance is extremely sensitive to the precise value of this threshold \cite{1975ApJ...201....1C}. Small variations in the threshold can lead to exponentially large changes in the final PBH abundance, making an accurate determination of the collapse threshold essential for reliable predictions of the PBH mass function. This generally requires relativistic numerical simulations, since the threshold for black hole formation depends strongly on the perturbation profile; see Ref.~\cite{Escriva:2021aeh} for a review. Fully relativistic numerical simulations allow one to follow the nonlinear evolution of large-amplitude perturbations and determine whether a given initial configuration disperses or collapses into a black hole. They also determine the resulting PBH mass, whose critical-scaling behaviour is crucial for characterising the low-mass tail of the PBH mass function~\cite{Niemeyer:1997mt}.

For these reasons, in this work we use fully relativistic numerical simulations to determine the critical conditions for collapse under our equation of state, following the numerical framework developed in Refs.~\cite{Escriva:2022bwe,Escriva:2022yaf}, and incorporate the resulting threshold variation into the peak-theory calculation of the PBH mass function following Ref.~\cite{Escriva:2023nzn}. The statistical abundance is then evaluated within the peak-theory framework \cite{Bardeen:1985tr}, as implemented for a time-dependent equation of state in Ref.~\cite{Escriva:2023nzn}. We restrict ourselves to the case of Gaussian fluctuations under the assumption of spherical symmetry.

In particular, we estimate the PBH abundance in the presence of thermal transitions using peak theory accounting for peaks of the Laplacian of the curvature fluctuation $\zeta$, namely $\Delta\zeta$ (see also Refs.~\cite{Pi:2024ert,Yoo:2020dkz} for the treatment in the case of a constant equation of state), adopting the representative peak profile employed in Ref.~\cite{Escriva:2023nzn},\footnote{Recently, it has been shown in Ref.~\cite{Escriva:2026hel} that the statistical dispersion of curvature profiles around the mean reference profile can be important when the power spectrum is sufficiently broad. We leave this effect beyond the scope of the present work and defer its investigation to future research.} We consider Gaussian curvature
fluctuations sourced by a nearly scale invariant power spectrum given by
\begin{equation}
 {\cal P}_{\zeta}(k)
 =
 A_{\zeta}
 \left(\frac{k}{k_0}\right)^{n_s-1},
 \qquad
 W(k/k_{\rm W})=\Theta(k_{\rm W}-k),
\end{equation}
where \(k_0\) is a fixed pivot scale defining the normalisation
\(A_\zeta\equiv {\cal P}_\zeta(k_0)\) for which we set $M_{H,0} = 1 M_{\odot}$ with $k_0 \approx 10^{6} \textrm{Mpc}^{-1}$, 
while \(k_{\rm W}\) denotes
the smoothing scale. The smoothing scale
\(k_{\rm W}\) varies across the calculation and is associated with
the horizon mass \(M_H\) of the perturbation at horizon crossing. The function \(W\) is a top-hat smoothing window in Fourier space. The spectral moments are defined as
\begin{equation}
\begin{aligned}
 \sigma_j^2
 &=
 \int\frac{{\rm d}k}{k}\,
 k^{2j}{\cal P}_{\zeta}(k)W^2(k/k_{\rm W})
 =
 \frac{A_{\zeta}k_0^{1-n_s}
 k_{\rm W}^{2j+n_s-1}}
 {2j+n_s-1},
 \\
 \gamma_3
 &=
 \frac{\sigma_3^2}{\sigma_2\sigma_4}.
\end{aligned}
\label{eq:spectral_moments}
\end{equation}
We parametrise the representative curvature profile by the amplitude
\begin{equation}
 \mu_2
 \equiv
 -\frac{2}{k_{\rm W}^2}
 \left.\Delta\zeta\right|_{r=0},
 \qquad
 k_\bullet^2
 =
 -\frac{\left.\Delta\Delta\zeta\right|_{r=0}}
 {\left.\Delta\zeta\right|_{r=0}},
 \label{eq:peak_parameters}
\end{equation}
where $\mu_2$ denotes the dimensionless amplitude associated with
the peak of $\Delta\zeta$ at the origin, while $k_\bullet$
characterizes its curvature scale.
In
the high-peak limit, the joint comoving number density is
\begin{align}
 n_{\rm pk}^{(\mu_2,k_\bullet)}
 ={}&
 \frac{2}{3^{3/2}(2\pi)^{3/2}}
 \frac{\sigma_2^3\sigma_4^2}
 {\sigma_1^4\sigma_3^3}
 \mu_2 k_\bullet\,f(\xi)
 P_1^{(3)}(\nu,\xi),
 \label{eq:joint_peak_density}\\
 \nu={}&
 \frac{\sigma_2}{\sigma_1^2}\mu_2,
 \qquad
 \xi=
 \frac{\sigma_2^2}{\sigma_1^2\sigma_4}
 \mu_2 k_\bullet^2 ,
 \nonumber
\end{align}
where \(f(\xi)\) is the BBKS peak-curvature function, whose explicit
expression can be found in Ref.~\cite{Escriva:2023nzn}, and
\begin{equation}
 P_1^{(3)}(\nu,\xi)
 =
 \frac{\exp\!\left[
 -\frac{1}{2}\left(
 \nu^2+\frac{(\xi-\gamma_3\nu)^2}{1-\gamma_3^2}
 \right)\right]}
 {2\pi\sqrt{1-\gamma_3^2}} .
\end{equation}


The representative curvature profile can be written as
\begin{equation}
    \zeta(r)=\mu_2\,g(x;\kappa),
    \qquad
    x\equiv \frac{r}{r_m},
    \qquad
    \kappa\equiv\frac{k_\bullet}{k_{\rm W}},
\end{equation}
in a radiation-dominated Universe, and at leading order in the gradient expansion \cite{Harada:2015yda}, the compaction function \cite{Shibata:1999zs} is given by $\mathcal{C}=\frac{2}{3}\left[1-(1+r\zeta')^2\right]$. We define $r_m$ as the characteristic scale of the curvature profile through the maximum of the linear compaction function, $\mathcal{C}_l=-\frac{4}{3}r\zeta'$, such that $\mathcal{C}_l'(r_m)=0$. We measure the threshold $\mathcal{C}_c$ relative to the radiation-dominated reference value. For the scale-invariant power spectrum, the
profile function is given by
\begin{align}
g(x;\kappa) =
\frac{6}{x^4\lambda^4}
\Bigg[
-12\kappa^2
+x^2\lambda^2\left(3-4\kappa^2\right) \qquad&
\nonumber\\
+ \, 4 \left(3\kappa^2-2\right)x\lambda\sin(x\lambda)
+8 \qquad&
\nonumber\\
+\left\{
12\kappa^2-8
+x^2\lambda^2\left(1-2\kappa^2\right)
\right\}
\cos(x\lambda)
\Bigg] \, , \quad&
\label{eq:curvature_profile}
\end{align}
with $\lambda\equiv k_{\rm W}r_m$. We extend this prescription to values around $n_s \approx 1$ by fixing
\begin{equation}
 k_\bullet=\kappa_t k_{\rm W},
 \qquad
 \kappa_t\simeq0.707,
 \qquad
 k_{\rm W}r_m\simeq4.16,
 \label{eq:fixed_profile}
\end{equation}
independently of $n_s$, since this choice yields the largest PBH abundance within the adopted representative-profile approximation. In addition, since the thermal variations considered here are smooth and relatively mild, the resulting shift in $\kappa_t$ is expected to be small, and we approximate it by its radiation-dominated value, $\kappa_t\simeq0.707$, as supported by the numerical analysis of \cite{Escriva:2023nzn}. All profile-dependent quantities are therefore evaluated using this fixed representative configuration. In
particular, the collapse threshold is taken to depend on the horizon
mass and the thermal equation of state, but not explicitly on the
spectral index $\mu_{2,c}(M_H;n_s)
 \simeq
 \mu_{2,c}(M_H)$. The spectral tilt is included only in the statistical abundance through
the moments in Eq.~\eqref{eq:spectral_moments}, and consequently
through \(\gamma_3\), \(\nu\), \(\xi\), and the peak number density.
Thus, varying \(n_s\) changes the relative statistical weight of peaks
at different smoothing scales while leaving the adopted profile and
its numerical collapse threshold unchanged. This constitutes one of the approximations adopted in our analysis and is expected to be reasonable because we consider only values of the spectral index close to $n_s = 1$. The threshold $\mu_{2,c}$ is then determined using fully relativistic numerical simulations in spherical symmetry with the \texttt{SPriBHoS} codes \cite{Escriva:2019sim,Escriva:2025eqc}.

We account for the critical collapse regime of PBH formation \cite{Niemeyer:1999ak} through:
\begin{equation}
 M_{\rm PBH}
 =
 K M_H
 \left[\mu_2-\mu_{2,c}(M_H)\right]^\gamma,
 \label{eq:critical_scaling}
\end{equation}
where $K$ is a profile-dependent constant, and $\gamma$ is the critical exponent associated with critical collapse. Owing to the computational complexity of determining the dependence of $K$ and $\gamma$ across the different thermal transitions considered here, we neglect their variation with the horizon mass $M_H$ and take the values for a radiation-dominated Universe with $\gamma \approx 0.356$ \cite{Evans:1994pj} (which is a universal quantity independent on the profile of the fluctuation) and $K \approx 5$, motivated by the results of Ref.~\cite{Escriva:2021pmf} for a range of perturbation profiles. This approximation is expected to capture the dominant effect on the PBH abundance, since its strongest sensitivity arises from the exponential dependence on the collapse threshold. At fixed \(k_{\rm W}\), the value of \(\mu_2\) associated with a PBH
of mass \(M_{\rm PBH}\), together with the corresponding Jacobian, is
\begin{equation}
\begin{aligned}
 \mu_2(M_{\rm PBH},M_H)
 &=
 \mu_{2,c}(M_H)
 +
 \left(\frac{M_{\rm PBH}}{K M_H}\right)^{1/\gamma},
 \\
 \left|
 \frac{{\rm d}\mu_2}{{\rm d}\ln M_{\rm PBH}}
 \right|
 &=
 \frac{\mu_2(M_{\rm PBH},M_H)-\mu_{2,c}(M_H)}
 {\gamma}.
\end{aligned}
\label{eq:critical_jacobian}
\end{equation}
We denote by
\begin{equation}
 \mathcal{N}_{\rm pk}(\mu_2;k_{\rm W})
 \equiv
 \int {\rm d}k_\bullet\,
 n_{\rm pk}^{(\mu_2,k_\bullet)}
 (\mu_2,k_\bullet;k_{\rm W})
 \label{eq:marginal_peak_density}
\end{equation}
the comoving peak number density per unit \(\mu_2\). In our calculation, the smoothing scales are restricted to those
associated with the range of horizon masses covered by our numerical collapse-threshold calculations, $10^{-4}\,M_\odot \leq M_H \leq 10^{11}\,M_\odot$. We do not extrapolate the numerically determined collapse threshold
outside this interval. The PBH
abundance per logarithmic mass interval is then
\begin{equation}
 \label{eq:conditional_PBH_mass_function}
\begin{aligned}
 f_{\rm PBH}(M_{\rm PBH}\vert k_{\rm W})
 =
 \frac{1}{\Omega_{\rm CDM}}
 \left(\frac{M_{\rm eq}}{M_H}\right)^{1/2}
 \frac{M_{\rm PBH}}{M_H} \quad&
 \\
 \times
 \frac{4\pi}{3}r_m^3\,
 \mathcal{N}_{\rm pk}
 \!\left(\mu_2(M_{\rm PBH},M_H);k_{\rm W}\right)&
 \\
 \times
 \left|
 \frac{{\rm d}\mu_2}
 {{\rm d}\ln M_{\rm PBH}}
 \right|& \,.
\end{aligned}
\end{equation}
Here, \(M_{\rm PBH}/M_H\) is the fraction of the horizon mass
incorporated into the PBH, while
\(\Omega_{\rm CDM}^{-1}(M_{\rm eq}/M_H)^{1/2}\) accounts for the
growth of the PBH energy fraction from formation during radiation
domination to matter--radiation equality.

Finally, the cloud-in-cloud
effect is treated by taking the envelope over the smoothing scales,
\begin{equation}
\begin{aligned}
 f_{\rm PBH}(M_{\rm PBH})
 &=
 \max_{k_{\rm W}}
 \left\{
 f_{\rm PBH}(M_{\rm PBH}\vert k_{\rm W})
 \right\},
 \\
 f_{\rm PBH}^{\rm tot}
 &=
 \int {\rm d}\ln M_{\rm PBH}\,
 f_{\rm PBH}(M_{\rm PBH}) .
\end{aligned}
\label{eq:mass_function_envelope}
\end{equation}
For each value of \(n_s\), the amplitude \(A_{\zeta}\) is adjusted
such that \(f_{\rm PBH}^{\rm tot}\) reproduces the desired total PBH
abundance i.e. the fraction of DM in PBHs.

\subsection{The Gravitational Waves from Primordial Black Hole Mergers} \label{subsec:method_GW}

Gravitational waves are generated by the accelerated motion of matter when it produces a time-varying quadrupole moment \cite{Einstein1918}. Therefore, coalescing PBH binaries emit GWs that may be detectable by current and future interferometers \cite{Sasaki_2018,LISACosmologyWorkingGroup:2023njw}. PBH binaries are commonly classified into two formation channels: the \textit{early}-Universe formation channel, in which nearby PBHs decouple from the Hubble expansion and become gravitationally bound \cite{Raidal2019}, and a \textit{late}-Universe channel, in which binaries form dynamically within virialised dark-matter halos, for example through close encounters accompanied by sufficient GW energy loss \cite{CLESSE2017105}. Both formation mechanisms can in principle coexist. In this work, we focus only on the \textit{late}-Universe gravitational-capture channel, while noting that the relative contributions of \textit{early}- and \textit{late}-formed binaries depend sensitively on the PBH abundance, spatial distribution, and subsequent dynamical evolution \cite{Ali-Haimoud2017}.

The aim of this subsection is to derive the predicted distribution of detectable PBH mergers in the $(\log _{10}M_B,q)$ plane, where $q \equiv M_A/M_B \leq 1$ and $M_B$ is the heavier component. Following the procedure of \cite{Bodeker:2020stj}, we compute the mass-dependent detection weight and normalise the resulting distribution to unity.
For each binary configuration $(M_A,M_B)$, we compute the relative contribution to the expected detection distribution. The intrinsic late-time merger weight is obtained by combining the mass dependence of the gravitational-capture rate with the number-weighted probabilities of drawing the two component masses from the adopted PBH mass function. Since only the shape of the final distribution is required, all mass-independent normalisation factors are omitted.
The remaining mass dependence of the intrinsic late-time merger distribution scales as~\cite{Kocsis:2017yty}:
\begin{equation}
\label{eq:late_merger_rate}
    \frac{{\rm d}^2\mathcal{R}_{\rm late}}
         {{\rm d}\ln M_A\,{\rm d}\ln M_B}
    \propto
    f_{\rm PBH}(M_A)\,f_{\rm PBH}(M_B)
    \frac{(M_A + M_B)^{10/7}}
         {(M_A M_B)^{5/7}}\,.
\end{equation}
The intrinsic merger distribution is then weighted by the detector sensitivity. We assume the 2025 Advanced LIGO Hanford noise curve at approximately O4b sensitivity \cite{Essick:2025zed,ligo_virgo_kagra_2026_gwtc5_psd} and restrict the signal to the frequency interval $50-2000\,\mathrm{Hz}$. 
For each binary, the detector range $r_{\rm det}(M_A,M_B)$ is evaluated from the chirp mass and the noise-weighted inspiral and merger contributions. The complete derivation and definitions of $r_{\rm det}$ and the related quantities are given in \cite{Magaraggia_2026}, which follows the idea described in \cite{Carr:2019kxo}. Assuming a Euclidean geometry, which is appropriate because LVK-like interferometers probe binaries predominantly in the local Universe where cosmological corrections are small, the accessible volume scales as
$V_{\rm det}(M_A,M_B)\propto r_{\rm det}^3(M_A,M_B)$, so that the detector-weighted contribution of each binary is proportional to the intrinsic late-time merger weight multiplied by $r_{\rm det}^3$. Binaries whose signal lies mostly outside the adopted detector band therefore receive a strongly suppressed weight.
The result is finally expressed in terms of the heavier component mass $M_B$ and the mass ratio $q$. The appropriate Jacobian is included when transforming the distribution from $(M_A,M_B)$ to $(\log_{10}M_B,q)$. The resulting two-dimensional distribution is normalised by its maximum value and therefore represents the relative detection weight across the $(\log_{10}M_B,q)$ plane. Observed gravitational-wave events can then be overlaid on the same plane to compare their component masses with the regions favoured by the adopted PBH mass function and detector selection function.

To evaluate the PBH merger rate, we assume only the \textit{late} merger rate given in 
Eq.~\eqref{eq:late_merger_rate}. PBH binaries can form at a very early stage when the 
Hubble radius grows to causally connect two PBHs~\cite{Raidal2019} (see 
also~\cite{Raidal:2024bmm} for an overview of formation mechanisms), or at a later stage 
in DM haloes~\cite{Clesse:2016vqa}. Whether \textit{early} or \textit{late} binaries 
dominate the merger rate has been a topic of discussion for the past decade; while the 
picture remains unclear, we justify our choice in this paragraph, as such a discussion is of 
primary importance for the constraints presented in Section~\ref{sec:discussion}. The 
analytical expression for the \textit{early} merger rate is given in~\cite{Kocsis:2017yty} 
(see also references therein); it was found to correlate with results from $N$-body 
simulations for monochromatic and log-normal PBH mass functions in~\cite{Raidal2019}. 
However, in~\cite{Kocsis:2017yty} the authors explicitly state that the expression holds 
only for PBH mass distributions with $M_{\rm PBH}^{\rm max}/M_{\rm PBH}^{\rm min} \lesssim 
10$. The merger rate expression is expected to depend sensitively on the breadth of the 
PBH distribution; see Sec.~4.1.6 of~\cite{LISACosmologyWorkingGroup:2023njw}.

A subsequent study of cosmological structure simulations with a lognormal PBH mass 
spectrum centered at $M_{\rm PBH} = 10~M_\odot$ found little to no correlation between the 
predicted merger rate and their numerical results~\cite{Delos:2024poq}. More recently, Aljaf \& Cholis \cite{Aljaf:2025dta} found that, although the contribution from binaries remaining isolated is suppressed at low redshift, dynamical interactions in haloes enhance the total merger rate in their models by $\approx 50\%$ at $z \lesssim 2$ relative to the assumption that all binaries remain unperturbed. Given that the latest studies 
suggest a suppression of the early merger rate\footnote{From a private discussion with 
S\'ebastien Clesse regarding the preliminary results of Simon Biot presented at the 
conferences \textit{Black Holes \& Cosmology 2026} and \textit{NEHOP 2026}: when 
considering a broad PBH mass spectrum, early isolated binaries are perturbed by the 
lighter surrounding PBHs. The lighter the PBH, the higher its number density and the 
greater the ability of the population to perturb early isolated binaries.} and that 
Ref.~\cite{Bodeker:2020stj} used only the \textit{late} merger rate, we follow the same 
assumption.

\section{Results}\label{sec:Results}

We present our results in this section, beginning with the EoS in 
Sec.~\ref{subsec:EoS_results}, followed by the PBH mass spectra in 
Sec.~\ref{subsec:pbh_mass_spectrum}, and finally a comparison of the merger rates with 
the Gravitational-Wave Transient Catalog (GWTC) in Sec.~\ref{subsec:merger_rate}.

\subsection{The cosmic equation of state}
\label{subsec:EoS_results}

\begin{figure*}[!ht]
    \centering
    \includegraphics[trim=1cm 0cm 1cm 0cm, clip, width=\linewidth]{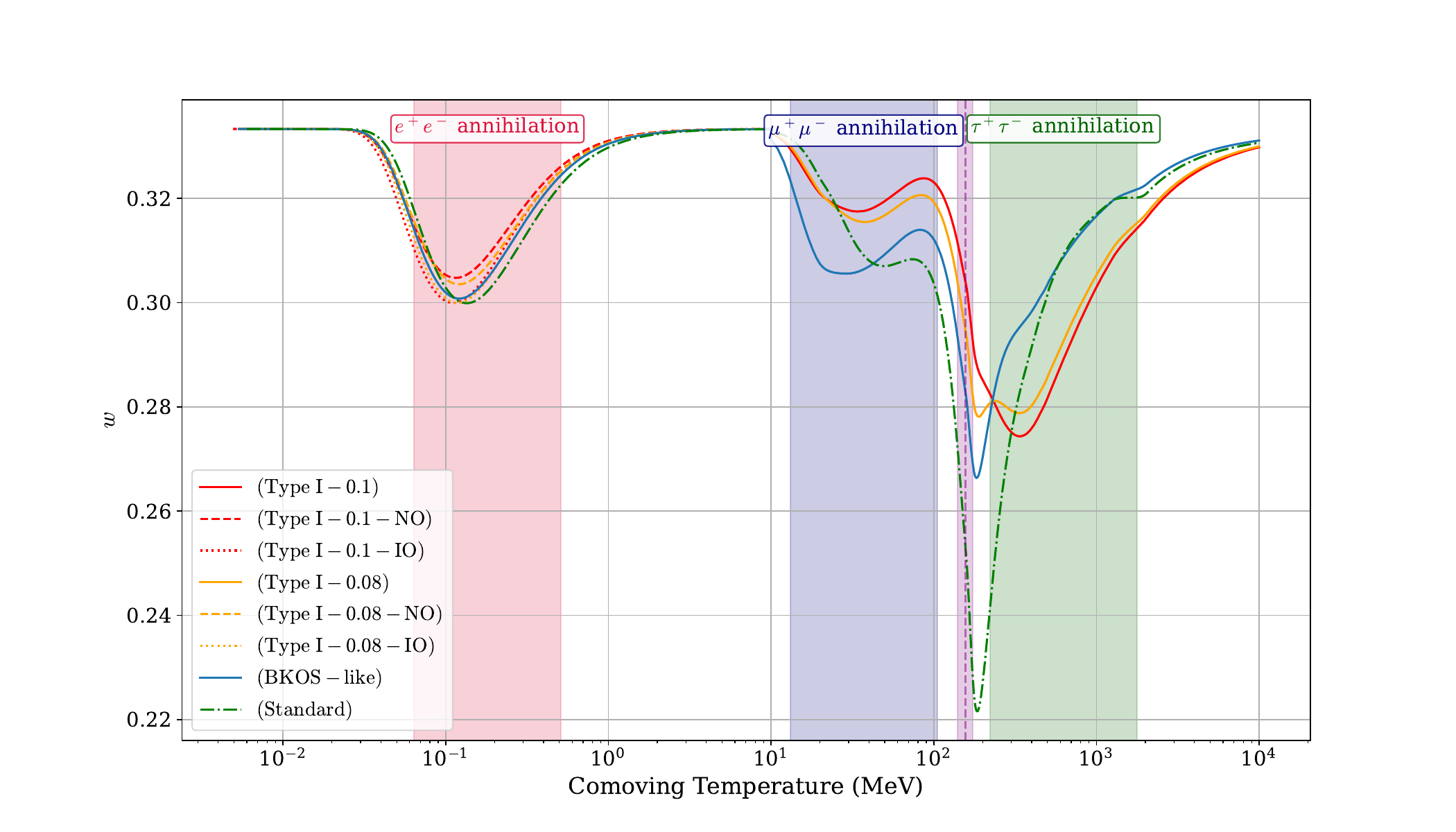}
    \caption{EoS of the various models considered; see Eqs.~\eqref{LAU:std}--\eqref{LAU:bodeker} for the corresponding LAU configurations. The purple shaded region around the pseudo-critical QCD temperature $T_c = 156.5 \, \mathrm{MeV}$ is purely illustrative. Results above $T = 10 \, \mathrm{MeV}$ are obtained with \texttt{CosmicEoS} and \texttt{Thermal-FIST}. Below this temperature, results are obtained with \texttt{NEVO} except for the $\mathrm{(Standard)}$ case, studied with \texttt{nudec\_BSM}; 
    see Secs.~\ref{subsec:EoS_QCD} and~\ref{subsec:nu_osc}. We use dashed (dotted) lines for normal (inverted) ordering of neutrino masses, denoted $\mathrm{NO}$ ($\mathrm{IO}$), except for the ${\rm (BKOS\text{-}like)}$ model where they are indistinguishable.
    Between $T = 1300~{\rm MeV}$ and $T = 2000~{\rm MeV}$, the thermodynamic 
    quantities have been interpolated to ensure a smooth transition between \texttt{CosmicEoS} 
    and the extrapolation parameters of~\cite{Bresciani:2025vxw}. Since no physical processes 
    are modelled in this intermediate region, the EoS parameter remains constant for the case 
    $\mu_B = 0$. The same interpolation method is applied to both the energy density 
    $\varepsilon$ and the pressure $P$; consequently, $w$ remains constant across all 
    trajectories. For non-vanishing $\mu_B$, the Taylor expansion makes the plateau in $w$ 
    less pronounced.}
    \label{fig:EoS}
\end{figure*}

We now turn to the impact of primordial asymmetries on the cosmic EoS. The 
introduction of chemical potentials changes the weight of each species in the primordial 
plasma: a large chemical potential $\mu_i$ tends to increase the contribution of the 
associated species $i$ (see Appendix \ref{app:species_contrib}). We show the EoS in Fig.~\ref{fig:EoS}. All models begin at 
$T = 10~{\rm GeV}$ with $w \sim 1/3$, the pure radiation value. The EoS quickly 
departs from this value as $\tau^+\tau^-$ annihilate and the QCD transition begins. The 
${\rm (Standard)}$ behaviour is well established~\cite{Borsanyi:2016ksw,Carr:2019kxo,
Gonin:2025uvc,Gonin:2026xhe}: we recognise the QCD dip, followed by the 
$\mu^+\mu^-$ and pion annihilation shoulder, and later, as 
$e^+e^-$ annihilate, another dip appears.

\paragraph{(BKOS-like) model.}
This model (blue line in Fig.~\ref{fig:EoS}) presents the same behaviour already found 
in~\cite{Bodeker:2020stj}, i.e.\ a mitigation of the QCD dip followed by an enhancement 
of the $\mu^+\mu^-$ and pion annihilation shoulder. Given our ability to describe the neutrino decoupling epoch, we show for the first time the 
$e^+e^-$ dip in the presence of LAU. At the level of the EoS, the results are almost indistinguishable between the two neutrino mass orderings, and we only show the NO one (see Appendix~\ref{app:details_NO_IO} for more details).
The dip appears slightly mitigated; the reason is that the $\nu_\alpha$ now carry $\mu_\alpha$, such that the neutrino/antineutrino sea takes a larger share of the thermodynamic weight. Even after asymmetry redistribution, the associated reheating maintains an increased neutrino contribution compared to the (Standard) case. Since neutrinos behave as pure radiation, their stronger thermodynamic weight pulls the EoS towards $w = 1/3$. The mitigation of the QCD dip can be interpreted in the same way, although at the QCD 
transition $\mu_B$ also affects the QCD sector directly. To disentangle these two effects 
we refer the reader to~\cite{Gonin:2026xhe}, where Fig.~5 shows the cosmic EoS with 
$\mu_B \neq 0$ and $\mu_{\nu_\alpha} = 0$. It is clear that $\mu_B \neq 0$ mitigates the QCD 
dip as well; thus the effects of the increased neutrino contribution and of $\mu_B \neq 0$ 
act in concert. Although the effect of $\mu_{\nu_\alpha} \neq 0$ is stronger than $\mu_B \neq 0$. 

The shoulder following the QCD dip is affected by $\mu_Q$, $\mu_\mu$, and 
$\mu_{\nu_\tau}$: a non-vanishing $\mu_Q$ impacts pion formation (see~\cite{Ferreira:2025zeu}), 
$\mu_\mu$ affects the $\mu^+\mu^-$ annihilation, and a non-vanishing $\mu_{\nu_\tau}$ 
renders the $\nu_\tau$ a significant thermodynamic contributor.

The minimum of the $e^+e^-$ dip appears shifted, but this is a numerical consequence of using the comoving temperature $T_\mathrm{cm}$ on the x-axis. We recall that during neutrino decoupling, we use $T_\mathrm{cm} \propto a^{-1}$ as a time variable in \texttt{NEVO}, and track the photon/electron/positron temperature $T_\gamma(T_\mathrm{cm})$. The location of the dip is set by the temperature at which $e^\pm$ become non-relativistic, i.e., $T_\gamma \sim m_e$. Because of the lepton asymmetry near-wash-out due to neutrino oscillations, all species get reheated~\cite{Froustey:2024mgf}. As a consequence, for a given $T_\mathrm{cm}$ after this asymmetry redistribution, $T_\gamma$ is \emph{higher} in the $\rm (BKOS-like)$ scenario than in the $\rm (Standard)$ one. That is why $T_\gamma \sim m_e$, and the associated dip in the EoS, is pushed to a \emph{lower} $T_\mathrm{cm}$ in Fig.~\ref{fig:EoS}.

\paragraph{(Type I) models.} The ${\rm (Type~I-0.1)}$ and ${\rm (Type~I-0.08)}$ models show noticeable features around the QCD transition. In these cases, the value of $\ell_\tau$ is large 
enough for the $\tau^+\tau^-$ annihilation to dominate the departure from $w = 1/3$. In 
the ${\rm (Type~I-0.1)}$ model, the $\nu_\tau$ contribution already dominates the 
thermodynamics at the QCD transition; as a result, the characteristic minimum at 
$T \sim T_c$ is replaced by an inflection point. We show the evolution of the individual 
contributions in Appendix~\ref{app:species_contrib}.

The LAU values of the ${\rm (Type~I-0.08)}$ model are not large enough to make 
the QCD minimum vanish, and the EoS instead presents a double-peak structure. Note that 
the $\rm (BKOS-like)$ model also shows the $\tau^+\tau^-$ annihilation as an 
inflection point, occurring around the $\tau$-driven minimum seen in the 
${\rm (Type~I-0.1)}$ and ${\rm (Type~I-0.08)}$ models.

Following the QCD transition, in the $\mu^+\mu^-$ annihilation area the EoS rises again, all LAU models present similar behaviour: the EoS peaks at larger values compared to the ${\rm (Standard)}$ case. 
This is again due to the $\nu_\tau$ pulling the EoS towards $w = 1/3$. Moreover, in the 
presence of LAU, the minimum of the shoulder is shifted to lower temperatures, because 
the $\mu^+\mu^-$ annihilation, while carrying asymmetries, dominates over the pion 
annihilation occurring in the same temperature range.

As the $\mu^+\mu^-$ pairs and pions annihilate, differences emerge between the 
$\rm (BKOS-like)$ model and the other LAU models. In the 
${\rm (Type~I-0.1)}$ and ${\rm (Type~I-0.08)}$ models, the $\nu_\tau$ 
contribution is too strong to allow the $\mu^+\mu^-$ annihilation to produce a 
departure from $w = 1/3$ as large as seen in the $\rm (BKOS-like)$ model.
This is one of our key results: large values of $\ell_\tau$ cause the EoS to depart from 
$w = 1/3$ prior to the QCD transition, but as the asymmetry is transferred to $\nu_\tau$, 
these neutrinos pull the EoS back towards $w = 1/3$ for the remainder of the radiation 
era.

Compared with the (Standard) case, the $e^+e^-$ dip shows two features. First, it is slightly shifted to lower comoving temperatures, as explained above when discussing the $\rm (BKOS-like)$ model. Second, we now see a difference between the two neutrino mass orderings. The NO dip is more dampened towards $1/3$. This is because in the IO case, the asymmetries are almost completely washed out by oscillations, at a temperature $\sim 4 \, \mathrm{MeV}$ for which neutrinos are not yet decoupled. This results in a strong reheating of all species together, leaving the balance of the different contributions to the EoS relatively unchanged with respect to the lepton-symmetric, (Standard) scenario. In the NO case, the equilibration is more gradual and some asymmetries remain, such that neutrinos retain a larger share of the energy density, hence driving $w$ towards $1/3$. We discuss these aspects in more detail in Appendix~\ref{app:details_NO_IO}.

In all cases, the mitigation of the $e^+e^-$ dip is less 
pronounced than the mitigation of the QCD dip. This is because, at the QCD transition, 
two effects act together: $\mu_B \neq 0$ directly affects the QCD sector, and the 
increased contribution of $\nu_\tau$ pulls the EoS towards $w = 1/3$. Additionally, the 
$\mu^\pm$ are still relativistic at this epoch and their associated chemical potential further 
increases their thermodynamic contribution. Only the large neutrino contribution is present at the $e^+e^-$ and neutrino decoupling epoch; the only massive and relativistic species at this stage are the $e^+e^-$ pairs, which carry no asymmetries there (see Fig.~\ref{fig:asym_evol}), hence the weaker mitigation.

\subsection{The Primordial Black Hole mass spectrum}
\label{subsec:pbh_mass_spectrum}

Having determined the EoS in our different scenarios, we can now calculate the collapse thresholds and resulting PBH mass spectra following the methodology outlined in Sec.~\ref{subsec:method_PBH_mass_function}. As a first step, we analyse how the critical amplitude depends on the horizon mass at the time of horizon crossing. This relation is essential because, for a fixed primordial power spectrum, the probability of PBH formation is exponentially sensitive to the value of the threshold as discussed before. Therefore, even moderate variations of the threshold induced by changes in the thermal history can lead to sizeable modifications of the predicted PBH abundance.

\begin{figure}
    \centering
    \includegraphics[width=\linewidth]{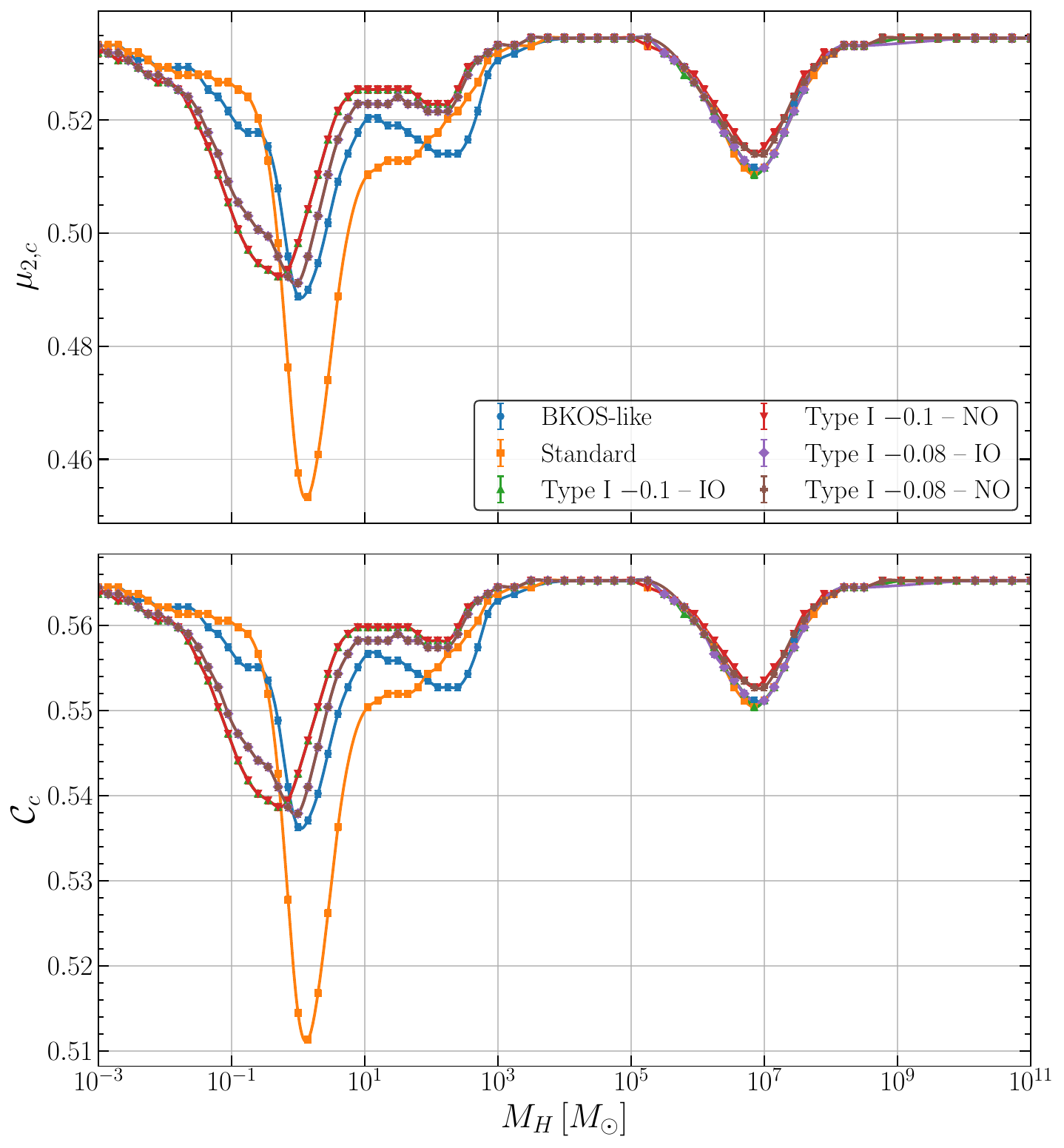}
    \caption{
    Threshold values for primordial black-hole formation as a function of the horizon mass ($M_H$) for different thermal histories of the early Universe EoS. The top panel shows the critical value of the curvature-profile amplitude, $\mu_{2,c}$, while the bottom panel shows the critical peak value of the compaction function, $\mathcal{C}_c$. The different curves correspond to the different cases considered in the analysis [Eqs.~\eqref{LAU:std}--\eqref{LAU:bodeker}]. Away from the regions where the equation of state varies appreciably, the thresholds approach an approximately constant radiation-dominated value. }
    \label{fig:thresholds_MH_EOS}
\end{figure}

Figure~\ref{fig:thresholds_MH_EOS} shows the threshold values as a function of the horizon mass $M_H(t_H)$ at the time $t_H$ when the fluctuation reenters the cosmological horizon, for the different equations of state considered in this work. The top panel displays the critical value of the curvature-profile amplitude,
$\mu_{2,c}$, whereas the bottom panel displays the critical peak value of the compaction function, $\mathcal{C}_c$. In the mass ranges where the EoS remains close to that of a pure radiation fluid, the threshold is approximately constant, as expected from the standard radiation-dominated result with value $\mu_{2,c} \approx 0.535$, $\mathcal{C}_c \approx 0.556$ obtained in \cite{Escriva:2023nzn}.
By contrast, distinct reductions in the collapse threshold appear at the mass scales associated with the softening of the EoS parameter $w$.
These reductions are visible in $\mu_{2,c}$, specifically, the minimum threshold is found around the solar-mass
scale, \(M_H(t_H)\approx M_\odot\). For a fixed EoS, both the magnitude of the threshold
reduction and the precise location of its minimum depend on the perturbation profile, as quantified through fully relativistic simulations of PBH formation at the QCD epoch in
Ref.~\cite{Escriva:2022bwe}.

The physical interpretation is straightforward. During the thermal transitions, the effective pressure support is temporarily reduced. As a consequence, overdense regions require a smaller initial amplitude to overcome pressure gradients and collapse gravitationally. This explains the dips observed in the threshold curves. The effect is particularly relevant for PBH phenomenology because the PBH abundance depends very sensitively on the threshold: a lower value of $\mu_{2,c}$ implies that a larger fraction of fluctuations
satisfies the collapse condition $\mu_2>\mu_{2,c}$, thereby
enhancing PBH formation at the corresponding horizon mass.
Therefore, the thermal history of the Universe can imprint characteristic features in the PBH mass spectrum, even when the primordial power spectrum itself is nearly scale-invariant as shown in previous studies with different scenarios of EoS \cite{Byrnes:2018clq,Carr:2019kxo,Escriva:2022bwe,Franciolini:2022QCD,Escriva:2022yaf,Escriva:2023nzn}.

\begin{figure*}[t]
    \centering
    \includegraphics[width=\linewidth]{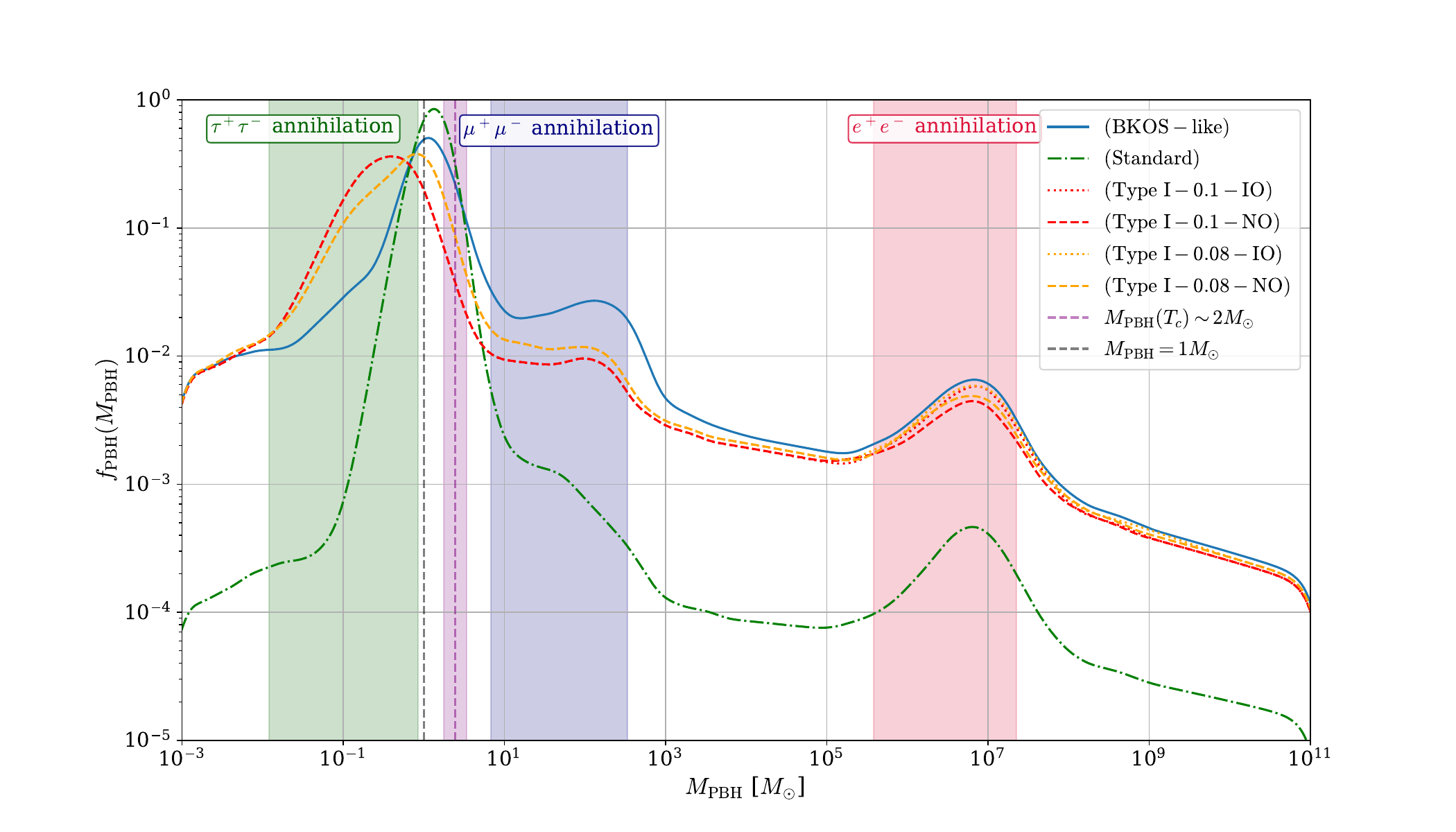}
    \caption{PBH mass spectra of the various models considered, normalised such that PBHs account for 
    all of the dark matter with a spectral index $n_s=0.97$ from CMB measurement, thus on much larger scales than the one considered for PBH formation; see Eqs.~\eqref{LAU:std}, \eqref{LAU:typeI_0.1}, \eqref{LAU:typeI_0.08}, \eqref{LAU:bodeker} for the corresponding LAU configurations. 
    ${\rm NO}$ and ${\rm IO}$ denote respectively the normal and inverted mass orderings of the 
    neutrinos. Prior to the $e^+e^-$ annihilation, the EoS for NO and IO are identical and the curves 
    overlap.
    Same colour code as Fig.~\ref{fig:EoS}.}
    \label{fig:pbh_spectra}
\end{figure*}

We plot the PBH mass spectra in Fig.~\ref{fig:pbh_spectra}. One can see how the dips in 
the EoS (Fig.~\ref{fig:EoS}) and in the threshold (Fig.~\ref{fig:thresholds_MH_EOS}) 
translate into peaks in the PBH mass spectra. In the presence of LAU, the QCD peak is 
less dominant in the distributions. The spectra are normalised using Eq.~\eqref{eq:mass_function_envelope}
with $f_{\rm PBH}^{\rm tot}$ the fraction of DM in PBHs for set to the same value all spectra; this typically leads to $A_\zeta \approx (2-3)\cdot10^{-3}$ depending on the case and $n_s$ considered, the 
remaining peaks increase their contribution by several orders of magnitude. As shown 
in~\cite{Bodeker:2020stj}, the presence of LAU actually reduces the total fraction of DM 
in PBHs for the same amplitude of the fluctuation spectrum. 

As seen with the $\rm (BKOS-like)$, ${\rm (Type~I-0.1)}$, and 
${\rm (Type~I-0.08)}$ models, the introduction of LAU flattens the distribution. 
Even though the QCD peak remains pronounced, we expect that such flat spectra would yield 
a highly non-trivial phenomenology. Given the current state of constraint evaluation, which 
often relies on monochromatic PBH mass distributions, it appears difficult to draw 
conclusions on which spectrum is favoured by observations.
For this reason, we allow 
ourselves to plot the spectra with $f_{\rm PBH}^{\rm tot} = 1$, i.e.\ PBHs accounting 
for all of the dark matter. These are extreme scenarios, most probably already excluded by 
current constraints~\cite{LISACosmologyWorkingGroup:2023njw,Byrnes:2025tji,Carr:2026hot}; nevertheless, they should be viewed as illustrative benchmarks to showcase the impact of the expected 
non-vanishing LAU.

Moreover, many free parameters can modify the distribution: the amplitude of the 
power spectrum of the curvature fluctuation $A_{\zeta}$ and the spectral index $n_s$ are unconstrained 
from observations on small scales 
(compared to the CMB anisotropy scales). As mentioned, changing the value of $A_{\zeta}$ modifies 
$f_{\rm PBH}^{\rm tot}$ without affecting the shape of the distribution. On the other 
hand, modifications of $n_s$ significantly change the shape of the spectrum; see 
Appendix~\ref{app:spectral_index}.
Additionally, one cannot exclude a running spectral 
index as in~\cite{Hasinger:2020ptw,Magaraggia_2026}, which could provide a 
distribution that evades the constraints while resolving cosmological conundrums.

With this in mind, we now return to the impact of LAU on the PBH mass spectrum. For the 
${\rm (Type~I-0.1)}$ and ${\rm (Type~I-0.08)}$ models, the large value of 
$\ell_\tau$ shifts the maximum of the `QCD peak'\footnote{As discussed in 
Section~\ref{subsec:EoS_results}, the dip in the EoS is actually attributed to $\tau^+\tau^-$ 
annihilation; `$\tau$+QCD peak' would therefore be more appropriate, but for consistency 
and easier comparison with the literature we retain the term `QCD peak.'} to smaller 
values of $M_{\rm PBH}$. The double-peak structure of ${\rm (Type~I-0.08)}$ 
visible in Fig.~\ref{fig:EoS} is no longer apparent in the mass spectrum, where it appears only as a mild inflection. This is not unexpected, since the threshold is determined by the full nonlinear collapse dynamics and does not simply trace the background EoS parameter $w$. In particular, pressure gradients, and hence the squared sound speed $c_s^2$, also play an important role and can
generate additional structure in the threshold, as demonstrated by fully relativistic simulations across the QCD transition in Ref.~\cite{Escriva:2022bwe}. The subsequent mapping from $M_H$ to $M_{\rm PBH}$, including critical collapse and the statistical weighting of fluctuations, further smooths these relatively small features, preventing them from appearing as distinct peaks in the final mass spectrum. 

The $M_{\rm PBH} \sim 10$--$10^2~M_\odot$ shoulder gains significant weight in the 
distribution for all LAU models, the most pronounced case being 
$\rm (BKOS-like)$: here $\ell_\mu$ increases the significance of the 
$\mu^+\mu^-$ annihilation, while $\ell_\tau$ is not large enough for $\nu_\tau$ to 
effectively compensate for the increased dip in the EoS. For the ${\rm (Type~I-0.1)}$ 
and ${\rm (Type~I-0.08)}$ models, the $\nu_\tau$ contribution helps to mitigate 
the dip in the EoS and hence the corresponding shoulder in the PBH mass distribution. 

Around the neutrino decoupling and $e^+e^-$ annihilation peak, the neutrino mass ordering 
does have an impact. The NO models show a stronger mitigation than the IO, see Appendix~\ref{app:details_NO_IO}.

Recently, some of us presented a new mechanism inducing mass growth of PBHs with $M_{\rm PBH} > 
10^3~M_\odot$ through absorption of the ambient neutrino radiation~\cite{Gonin:2026pqv}. 
Although the effect relies on established physical processes and cannot be prevented if a population of PBHs with $M_{\rm PBH} > 10^3~M_\odot$ forms around the epoch of neutrino decoupling, we do not take it into account in the present paper.\footnote{The models with $f_{\rm PBH}^{\rm tot}=1$ at formation, appear to be forbidden because the neutrino absorption would let $f_{\rm PBH}^{\rm tot}>1$, i.e. an excess of DM.} The process depends sensitively on the collapse fraction $\gamma = M_{\rm PBH}(T_i)/M_H(T_i)$,\footnote{Different from the critical exponent from \eqref{eq:critical_scaling}, in the literature the same symbol is often used to describe these two different quantities.} where $T_i$ is the temperature at formation, $M_H(T_i)$ the horizon mass at horizon crossing, and $M_{\rm PBH}(T_i)$ the PBH mass at formation. To keep the present paper concise, we avoid introducing an additional free parameter. Moreover, the evaluation of neutrino absorption in the presence of LAU requires careful treatment; we defer this to future studies.

\subsection{The Gravitational Waves from Primordial Black Holes}\label{subsec:merger_rate}

One exciting observational channel for PBHs is through gravitational wave observations; 
the growing number of binary black hole (BBH) merger detections is ushering in an era of 
statistical studies of BBH mergers. PBHs have already been proposed as candidates for some 
BBH observations~\cite{Clesse:2020ghq,Franciolini:2022QCD,DeLuca:2025fln}, while 
Ref.~\cite{Bodeker:2020stj} suggested that LAU can help reconcile the BBH population with 
the PBH merger probability density. With this in mind, we attempt not to exclude models, 
but rather to identify the parameters that could explain the current observations in the 
GWTC-5~\cite{LIGOScientific:2026pop,
LIGOScientific:2026wfs}. The aim of this section is to highlight tendencies in models 
that could inform the mapping of relevant configurations for gravitational wave 
observations; these are the models that should be targeted by accurate constraining 
methods.

The impact of the QCD transition on PBH merger phenomenology was also studied in Ref.~\cite{Escriva:2022bwe}, where fully relativistic collapse thresholds for the time-dependent QCD equation of state were incorporated into the PBH mass function and compared with GWTC-3~\cite{KAGRA:2021vkt}. That analysis found that the resulting merger distributions could account for part of the $30$--$50\,M_\odot$ population and exceptional events such as GW190814 and GW190521, while the absence of a secondary $8$--$15\,M_\odot$ feature pointed towards the possibility of a mixed primordial and astrophysical population. The present work extends this line of investigation by including lepton asymmetries, neutrino oscillations, peak theory, and the updated GWTC-5 catalogue.

\begin{figure*}[!h]
    \centering
    \begin{subfigure}[b]{0.33\textwidth}
        \centering
        \includegraphics[width=\textwidth]{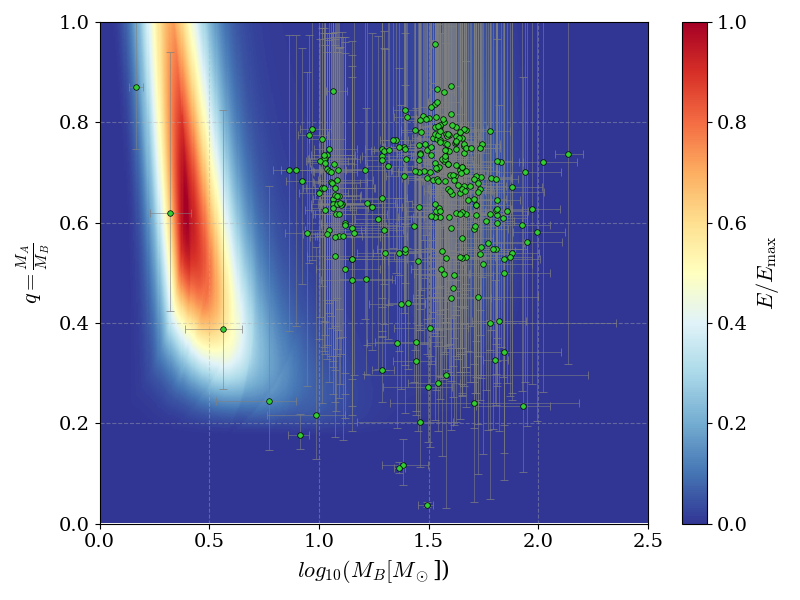}
        \caption{${\rm (Standard)}, ~n_s=0.94$}
    \end{subfigure}%
    \begin{subfigure}[b]{0.33\textwidth}
        \centering
        \includegraphics[width=\textwidth]{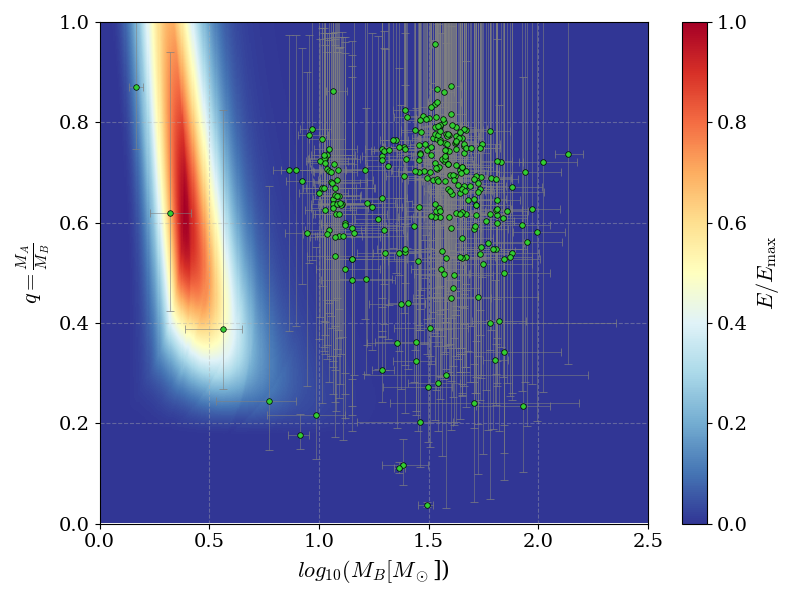}
        \caption{${\rm (Standard)}, ~n_s=0.97$}
    \end{subfigure}%
    \begin{subfigure}[b]{0.33\textwidth}
        \centering
        \includegraphics[width=\textwidth]{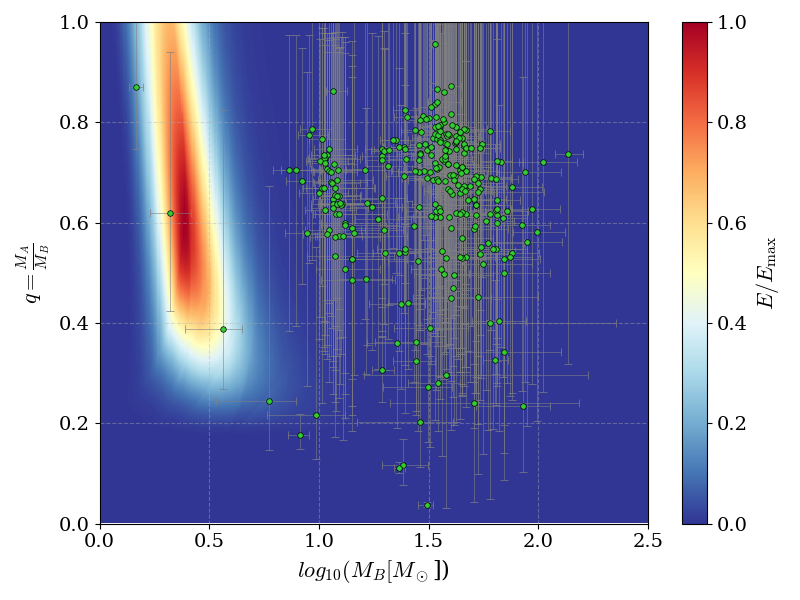}
        \caption{${\rm (Standard)}, ~n_s=0.99$}
    \end{subfigure}
    \begin{subfigure}[b]{0.33\textwidth}
        \centering
        \includegraphics[width=\textwidth]{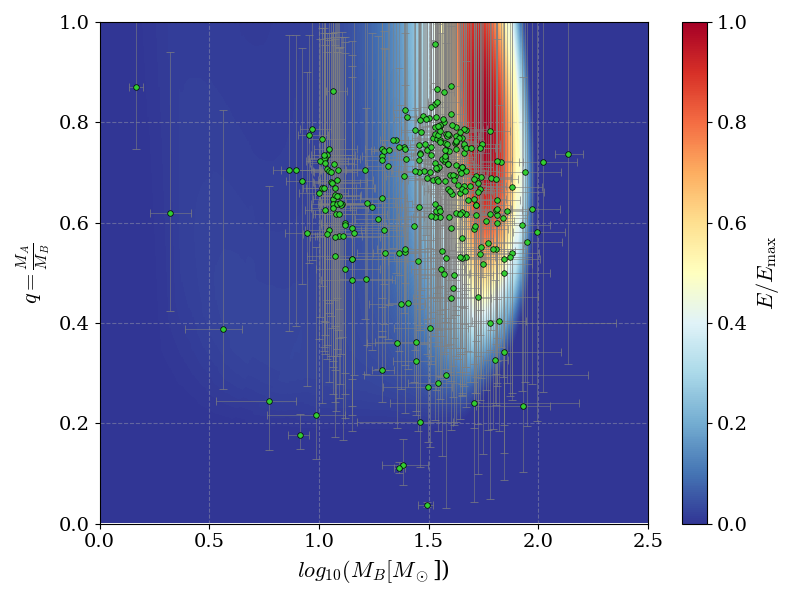}
        \caption{${\rm (BKOS\text{-}like)}, ~n_s=0.94$}
    \end{subfigure}%
    \begin{subfigure}[b]{0.33\textwidth}
        \centering
        \includegraphics[width=\textwidth]{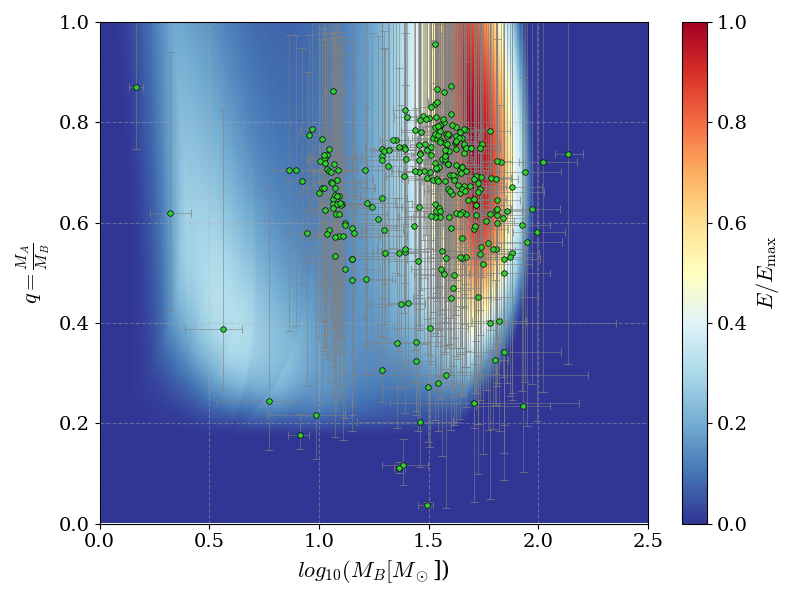}
        \caption{${\rm (BKOS\text{-}like)}, ~n_s=0.97$}
    \end{subfigure}%
    \begin{subfigure}[b]{0.33\textwidth}
        \centering
        \includegraphics[width=\textwidth]{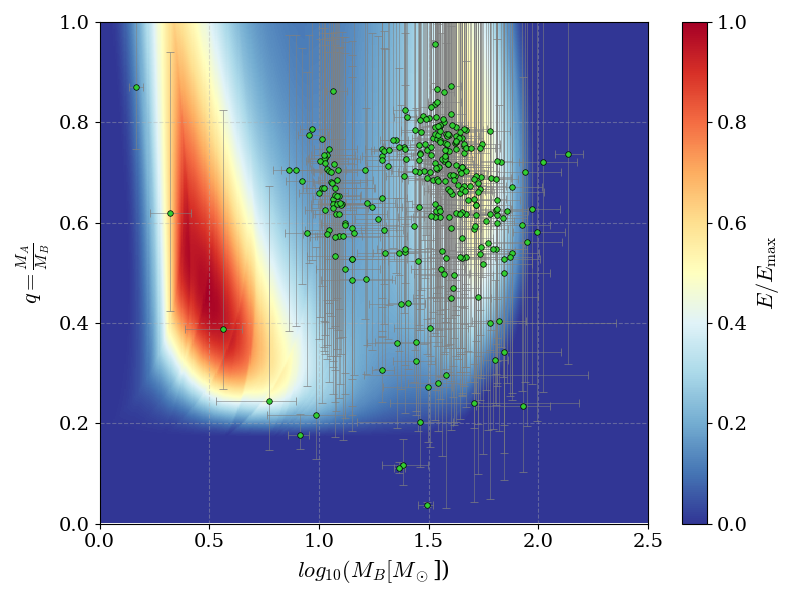}
        \caption{${\rm (BKOS\text{-}like)}, ~n_s=0.99$}
    \end{subfigure}
    \begin{subfigure}[b]{0.33\textwidth}
        \centering
        \includegraphics[width=\textwidth]{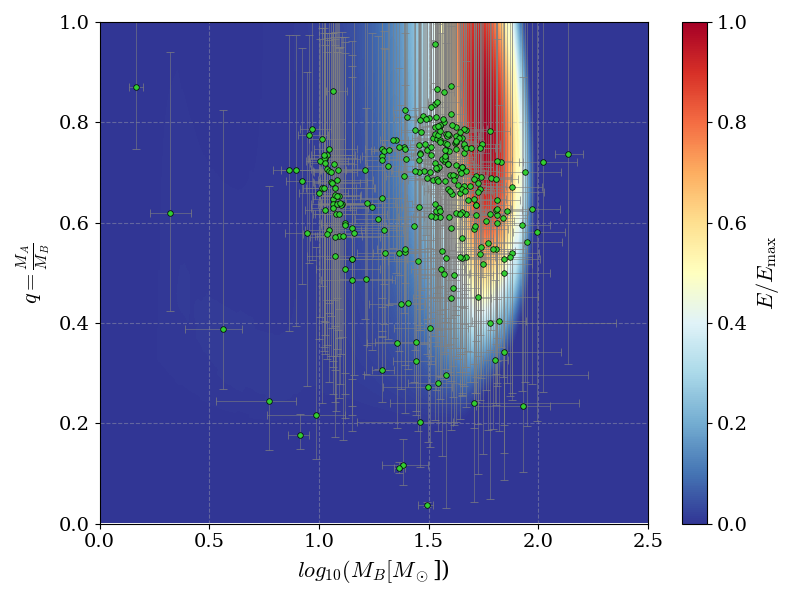}
        \caption{${\rm (Type~I-0.08)}, ~n_s=0.94$}
    \end{subfigure}%
    \begin{subfigure}[b]{0.33\textwidth}
        \centering
        \includegraphics[width=\textwidth]{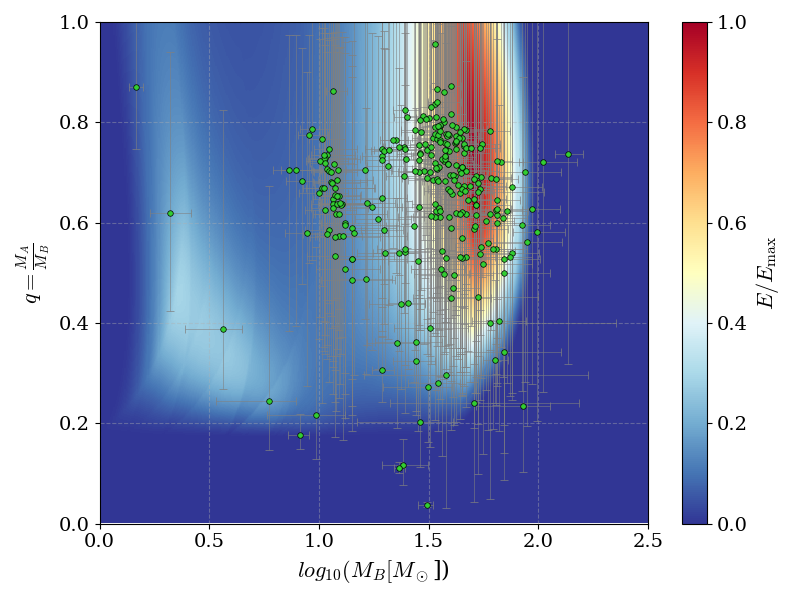}
        \caption{${\rm (Type~I-0.08)}, ~n_s=0.97$}
    \end{subfigure}%
    \begin{subfigure}[b]{0.33\textwidth}
        \centering
        \includegraphics[width=\textwidth]{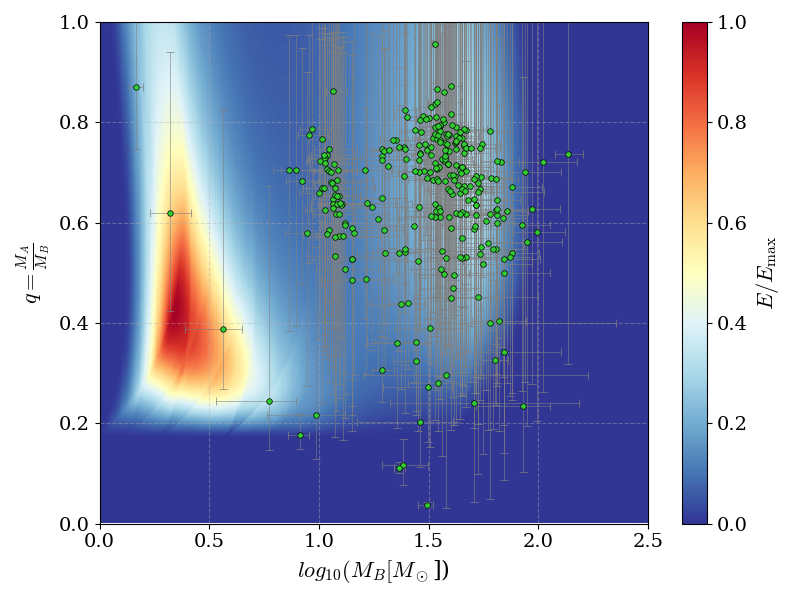}
        \caption{${\rm (Type~I-0.08)}, ~n_s=0.99$}
    \end{subfigure}
    \begin{subfigure}[b]{0.33\textwidth}
        \centering
        \includegraphics[width=\textwidth]{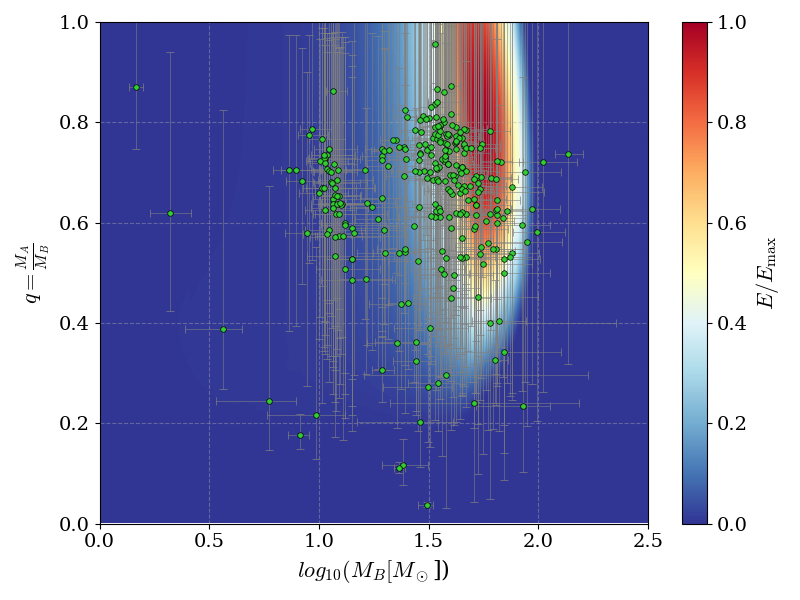}
        \caption{${\rm (Type~I-0.1)}, ~n_s=0.94$}
    \end{subfigure}%
    \begin{subfigure}[b]{0.33\textwidth}
        \centering
        \includegraphics[width=\textwidth]{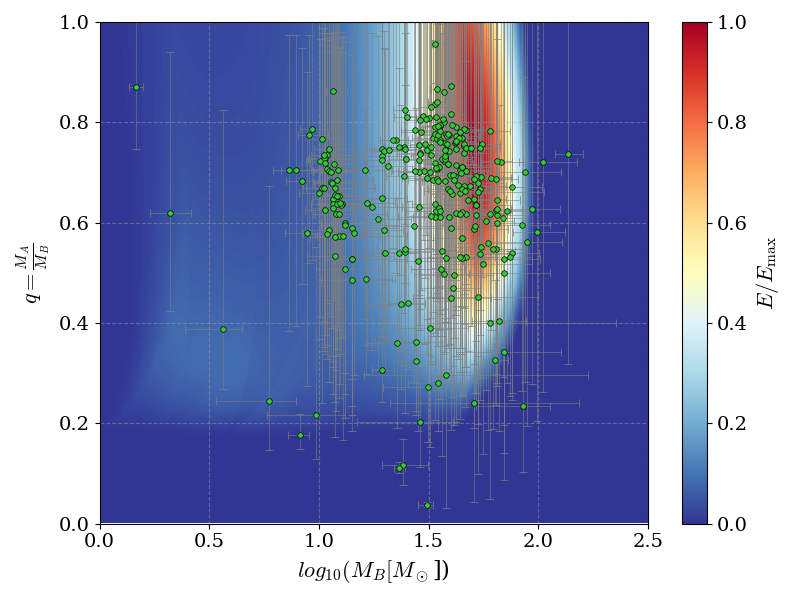}
        \caption{${\rm (Type~I-0.1)}, ~n_s=0.97$}
    \end{subfigure}%
    \begin{subfigure}[b]{0.33\textwidth}
        \centering
        \includegraphics[width=\textwidth]{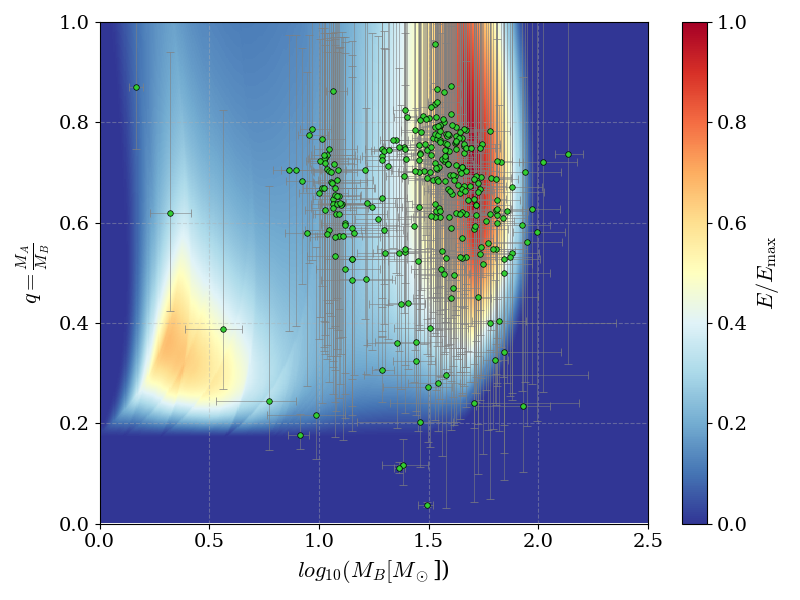}
        \caption{${\rm (Type~I-0.1)}, ~n_s=0.99$}
    \end{subfigure}
    \caption{Probability density of PBH merger detections with masses $M_B>M_A$ by LVK collaboration, with the parameter space spanned by $M_B$ and $q=M_A/M_B$. The green dots show the GWTC-5 events mapped onto the parameter space. 
    }
    \label{fig:grid}
\end{figure*}

To this end, we follow a procedure similar to that of Ref.~\cite{Bodeker:2020stj}, 
overlaying the observed events from the cumulative GWTC-5 onto the detection likelihood 
derived from the PBH merger rate constructed from our distributions; see Section~\ref{subsec:method_GW}.
The results are shown in Fig.~\ref{fig:grid}, where the 
LAU is fixed along rows and the spectral index $n_s$ varies across columns. Note that since we 
consider only frequencies in the range $50$--$2000~{\rm Hz}$, only PBHs around the QCD 
peak contribute; mergers from supermassive PBHs associated with the $e^+e^-$ annihilation 
peak do not contribute in this frequency band.

Ref.~\cite{Bodeker:2020stj} already found that certain LAU configurations are preferred 
to explain the observed GW events; the results of the present study corroborate this 
finding. However, in light of GWTC-5, there appear to be two main BBH groups: the most 
populated, with $\log_{10}(M_B/M_\odot) \sim 1.5$--$2$ ($M_B$ being the mass of the 
heavier binary component) and $q \sim 0.4$--$0.8$; and a secondary group at 
$\log_{10}(M_B/M_\odot) \sim 1$--$1.5$ and $q \sim 0.4$--$0.8$. Other events are 
distributed outside these two groups; some of the low-mass events appear to align with the 
${\rm (Standard)}$ case and its prominent QCD peak. However, this model cannot explain 
the bulk of the BBH population for the range of spectral indexes considered. 

A look at the changes brought by variations of $n_s$ shows that $n_s \sim 0.99$, where 
the QCD peak is enhanced (see Appendix~\ref{app:spectral_index}), concentrates the 
probability distribution towards low-mass mergers. This effect is less pronounced for 
${\rm (Type~I\text{-}0.1)}$; this case would be compelling if more events were to 
populate the secondary group at low mass. It is clear that $n_s \gtrsim 1$ would render 
the PBH population from thermal history unable to explain the BBH mergers, a result that 
also appeared in~\cite{Franciolini:2022QCD}.

$n_s = 0.94$ strengthens the main probability distribution for the LAU models while 
suppressing its spread. The secondary group of observed events then cannot be explained by PBH mergers. One might 
argue that the observed BBH population is likely to be a mixture of astrophysical black 
holes and PBHs, and that the secondary group could signal an astrophysical origin; 
however, the merger rate does not rest on sufficiently strong theoretical ground to draw 
such conclusions. Some inferred BBH parameters could help break the degeneracy between 
astrophysical and primordial origins~\cite{Franciolini:2021xbq}.
Nevertheless, models with $n_s = 0.94$ have the $e^+e^-$ annihilation peak as the 
dominant contribution; without a cut-off in the fluctuation spectrum at the corresponding 
scales, the dominance of heavy PBHs makes these scenarios difficult to reconcile with 
current constraints~\cite{DeLuca:2025nao}.

We find that $n_s \lesssim 0.99$ combined with LAU provides the best primordial explanation of GWTC-5 events. Note that the analysis provided in this section is independent of the value of 
$f_{\rm PBH}^{\rm tot}$.
Of course, the procedure outlined in Section~\ref{subsec:method_GW} is model-dependent, and applying it to such non-trivial PBH mass spectra may venture outside the domain of 
validity of the model. This is the main reason why we refrain from quantitative 
observational evaluation; the analysis from this section should be viewed as indicative 
rather than conclusive. In the following section we provide a brief summary of the 
constraint mechanisms likely to apply to the spectra presented here.

The results from this section may be compared with those of~\cite{Franciolini:2022QCD}, 
as they identify preferred values of $n_s$; however, there are several key differences in 
both the ambitions and assumptions of the two studies. First, \cite{Franciolini:2022QCD} 
uses the \textit{early} merger rate; as mentioned in Sec.~\ref{subsec:method_GW}, the analytical formula 
of~\cite{Kocsis:2017yty} should not be applied to PBH mass spectra spanning more than two 
orders of magnitude, and the authors acknowledge this theoretical uncertainty. Both studies utilise peak theory, albeit in different formulations: Franciolini {\it et al.} \cite{Franciolini:2022QCD} use high-peak statistics of the smoothed linear density contrast, while we follow the formulation based on peaks of the Laplacian of the curvature perturbation described in Sec.~\ref{subsec:method_PBH_mass_function}. They reconstruct an 
\textit{ab initio} PBH mass spectrum from the QCD EoS without LAU by performing Bayesian 
inference over GWTC-3~\cite{KAGRA:2021vkt}, ultimately inferring the associated 
fluctuation spectrum and inflationary model. The scope of~\cite{Franciolini:2022QCD} and the present study differ, as we focus rather 
on the description of the primordial plasma. Nevertheless, their results remain relevant in our context: they find that smaller values $n_s < 1$ are favoured, the reason being that decreasing $n_s$ increases the weight of PBHs from the $\mu^+\mu^-$ and pion annihilation era. LAU can naturally produce such behaviour without a significant departure from scale invariance at $n_s = 1$. They use their results to constrain the PBH abundance as a dark matter candidate in the QCD transition mass range. We discuss constraints in the following section.

\section{Discussion}\label{sec:discussion}

Several PBH mass functions and corresponding merger-rate predictions obtained for nonzero lepton asymmetries provide a better fit to the gravitational-wave data than the zero-asymmetry benchmark. This inevitably raises the question of their compatibility with existing observational constraints: the resulting mass functions na{\"i}vely seem to violate a number of constraints stemming from GW, microlensing, accretion, and $\mu$-distortions. This apparent tension does not, however, necessarily imply that the model is fully excluded. 

We acknowledge that existing constraints may challenge or exclude the possibility that PBHs in the mass range considered here account for all of the dark matter, i.e.~$f_{\rm PBH}^{\rm tot} = 1$.
Importantly, our analysis and the effects on the PBH mass spectrum emerging from LAU are independent of the value of $f_{\rm PBH}$. 
In any case, the asteroidal mass range remains an entirely open window which could accommodate most or even the entirety of the dark matter~\cite{
LISACosmologyWorkingGroup:2023njw}.
In fact, ultra-slow-roll inflation models can generate doubly-peaked fluctuation spectra, with peaks at the asteroid horizon mass scale and at 
the QCD transition scale~\cite{Franciolini:2022Engineering,Franciolini:2022QCD}. Even if the considered mass range cannot dominate the DM content, PBHs could still provide a significant contribution to GW observations, as explicitly demonstrated for PBHs formed around the QCD epoch in Ref.~\cite{Escriva:2022bwe}.

First of all, the commonly displayed PBH exclusion curves should not be applied uncritically to the broad, multi-peaked spectra obtained in this work. Most constraints were derived for monochromatic mass functions and depend on assumptions about the spatial and velocity distributions of PBHs, their clustering, the Galactic halo model, source populations and survey efficiencies. Although prescriptions exist for recasting monochromatic bounds for extended distributions \cite{Kuhnel:2017pwq, Carr:2017jsz, Bellomo:2017zsr}, they retain the astrophysical and observational assumptions of the underlying analyses. Their applicability must therefore be assessed separately for each observable, particularly when the response is nonlinear in the mass function or depends on PBH clustering and evolution. Moreover, these recast bounds do not replace a likelihood analysis tailored to the multi-peaked spectra considered here.

These uncertainties are especially apparent for microlensing towards the Magellanic Clouds. The strong limits reported by OGLE \cite{Mroz:2024mse, Mroz:2024wag} have been critically re-examined by Hawkins \& Garc{\'i}a-Bellido \cite{Garcia-Bellido:2024yaz, Hawkins:2025mlo}. They emphasise that OGLE recovered far fewer events than those expected from known stellar populations under the adopted modelling, thereby raising questions about the inferred detection efficiency. They also identify substantial sensitivity to the photometric passbands, crowding and blending, event-selection criteria, self-lensing contribution and assumed halo model. Furthermore, the OGLE result is in tension with the earlier MACHO event excess~\cite{MACHO:2000qbb}, interpreted as evidence for a compact halo population. Separately, Garc{\'i}a-Bellido \& Hawkins reanalysed the MACHO and EROS-2 results using a Galactic halo model informed by Gaia DR3 and found that an extended thermal-history PBH mass function can remain compatible with PBHs comprising the entirety of the dark matter \cite{Garcia-Bellido:2024yaz}.

Similar qualifications apply elsewhere. Supernova microlensing limits depend on the source structure and spatial PBH distribution \cite{Garcia-Bellido:2017imq, DES:2024ffp}, while accretion constraints depend on uncertain gas dynamics, radiative efficiencies and feedback, as well as the relative velocities of PBHs and the ambient gas \cite{Ali-Haimoud:2016mbv, Agius:2024ecw}. At the highest masses, CMB $\mu$-distortion constraints, derived from enhanced small-scale curvature perturbations \cite{Sharma:2024img}, depend on the assumed statistics of the primordial perturbations and on the mapping from the curvature profiles to the PBH abundance \cite{Byrnes:2024Mu,Yang:2025Compaction}. Recently, it has been demonstrated \cite{Dave:2026gsd} that a broad class of large non-Gaussianities can even entirely remove all $\mu$-distortion bounds.
Scalar-induced gravitational waves, sourced by second-order primordial scalar perturbations, constitute a complementary probe of the curvature power spectrum underlying PBH formation~\cite{Domenech:2021ztg}. This channel is also relevant to the (sub)solar-mass range through pulsar timing arrays. At lower masses, Blas {\it et al.}~\cite{Blas:2026xws} investigate prospective lunar and satellite laser-ranging searches in the $\mu{\rm Hz}$ band, which can probe perturbations corresponding to planetary-mass PBHs.

Ref.~\cite{Pritchard:2024vix} constrained the fluctuation power spectrum through various observational channels using a scenario similar to the one presented here. They used the 
standard QCD phase transition to bound the parameters of the fluctuation power spectrum, but did not assume a scale-invariant spectrum as in the present work, and did not account for LAU. As 
already found in Ref.~\cite{Bodeker:2020stj}, for a fixed value of $A_\zeta$, LAU reduces the 
PBH abundance from the QCD transition and would require to rederive the bounds.

We introduced our assumptions regarding the merger rate calculation in Sec.~\ref{subsec:method_GW}, in particular, the fact that we focus on the \emph{late} merger rate. The recent limits of Ref.~\cite{Andres-Carcasona:2026avd} based on the O4a run employ early-Universe two- \& three-body PBH binary-formation channels and neglect late formation. Although that work explicitly notes that the QCD transition can generate additional structure in the PBH mass function, its quantitative bounds were obtained for monochromatic and log-normal mass functions rather than for the broad, multi-peaked thermal-history spectra considered here. Applicability to the present spectra therefore requires a dedicated likelihood analysis. The same analysis also derives bounds that are agnostic about the astrophysical binary black hole population; by contrast, in joint primordial--astrophysical population analyses, the inferred PBH contribution can depend sensitively on the astrophysical models included~\cite{Franciolini:2022Subpopulation}. Specific measurements of binary black hole masses, mass ratios and spins in GWTC-5~\cite{LIGOScientific:2026pop} may help constrain the astrophysical population and thereby reduce the latter uncertainty.

Existing PBH abundance constraints should therefore not (at least, yet) be regarded as a threat to the present scenario. Nevertheless, the quantitative viability of the benchmark spectra must ultimately be established by evaluating the relevant likelihoods for the specific mass functions predicted here. A definitive assessment would require a dedicated analysis incorporating the full extended mass spectrum \cite{Carr:2017jsz, Bellomo:2017zsr}, the {\it unavoidable} Poisson fluctuations associated with PBH discreteness, their subsequent gravitational evolution and clustering \cite{ClusteringXXX}, and any additional primordial non-Poisson correlations \cite{Choi:2025eqn}.

It should furthermore be stressed that existing observations provide not only upper limits on the PBH abundance, but also a growing number of conundra that may be regarded as positive indications for PBHs \cite{Carr:2023tpt}. Perhaps the most striking example is Abell~2744-QSO1, a gravitationally lensed Little Red Dot at ($z = 7.04$), which contains a massive black hole in a dynamically light and nearly pristine host with a metallicity below $10^{-2}\,Z_{\odot}$ \cite{Maiolino:2025tih}. The coexistence of such a massive black hole with so little stellar mass and chemical enrichment is difficult to explain through conventional stellar-remnant, direct-collapse or super-Eddington-growth scenarios. By contrast, a massive PBH would naturally precede both the formation of the host galaxy and its chemical enrichment. Indeed, cosmological simulations starting from a massive PBH seed have recently been shown to reproduce the low metallicity, extreme black-hole-to-stellar-mass ratio and comparatively weak accretion inferred for QSO1 \cite{Zhang:2025oyl}. Such seeds may arise naturally from the high-mass peaks of thermal-history PBH mass functions, including the one associated with electron--positron annihilation.

A particularly remarkable set of further conundra concerns the unexpectedly early appearance of luminous sources observed by JWST, such as JADES-GS-z14-0, whose redshift was refined to $z_{\rm spec} = 14.1796 \pm 0.0007$ \cite{Carniani:2024zaf}, and MoM-z14 at $z_{\rm spec} = 14.44 \pm 0.02$ \cite{Naidu:2025xfo}. Beyond this confirmed frontier, the MIDIS+NGDEEP observations have yielded nine photometrically selected candidates spanning $16 < z < 25$, from which UV luminosity functions have been inferred at $z \sim 17$ and even $z \sim 25$ \cite{Perez-Gonzalez:2025bqr}. If confirmed spectroscopically, their inferred abundance would substantially increase the tension with conventional galaxy-formation models. Intriguingly, the corresponding UV luminosity density can be reproduced by PBHs with masses $M_{\rm PBH} = 10^{4-5}\,M_{\odot}$, residing in low-mass haloes and accreting at a moderate fraction of their Eddington luminosity before the onset of significant star formation \cite{Matteri:2025vnv}. At an even more speculative frontier, the extreme F356W-dropout Capotauro admits an extragalactic fit at $z \sim 32$, merely about $100\,{\rm Myr}$ after the Big Bang \cite{2026A&A...706A.364G}. Spectroscopic confirmation, particularly together with evidence for accretion-powered emission, would constitute compelling evidence for a non-stellar and possibly primordial origin. A growing number of papers (see, e.g., Refs.~\cite{Dayal:2024zwq, DeLuca:2025nao, Zhang:2025oyl}) consider a possible primordial origin of the Little Red Dots.

Further observational conundra include the aforementioned microlensing event excess towards the Magellanic Clouds reported by MACHO~\cite{MACHO:2000qbb}; long-duration Galactic-bulge microlensing events consistent with black-hole lenses \cite{Wyrzykowski:2019jyg, Niikura:2019kqi}; quasar microlensing suggestive of a cosmologically distributed population of compact objects~\cite{Hawkins:2020zie}; the masses, spins and merger rates of parts of the LVK binary-black-hole population~\cite{KAGRA:2021vkt, KAGRA:2021duu, Clesse:2017bsw}; the excess source-subtracted near-infrared CIB anisotropies discovered in Ref.~\cite{Kashlinsky:2005}, which may
be generated by the enhanced abundance of early halos arising from the
Poissonian white-noise contribution of PBH dark matter to the
small-scale power spectrum \cite{ Kashlinsky:2016sdv}, first proposed by in Ref.~\cite{Meszaros:1975} before inflationary theory appeared, while their coherence with the unresolved
CXB uncovered in Ref.~\cite{Cappelluti:2013} and confirmed in Ref.~\cite{Mitchell-Wynne:2016, Cappelluti:2017, Li:2018} independently indicates a high abundance of accreting black holes
amongst the CIB sources \cite{Helgason:2014, Hasinger:2020, Cappelluti:2022} (see review \cite{Kashlinsky:2018mnu}); and the dynamical properties of ultra-faint dwarf galaxies \cite{Simon:2019nxf, Clesse:2017bsw}. Taken together, these conundra constitute an increasingly broad and mutually complementary body of observational evidence motivating a PBH interpretation.

Additionally, we would like to point out that the mass spectra in Fig.~\ref{fig:pbh_spectra} have been normalised to $f_{\rm PBH}^{\rm tot} = 1$ as an illustrative benchmark chosen to expose the effects of the lepton asymmetries and the spectral index. Our principal result concerns the resulting redistribution of PBHs among the different thermal-history peaks, rather than the particular normalisation. Within the adopted parametrisation, reducing $f_{\rm PBH}^{\rm tot}$ changes the overall abundance without erasing these characteristic features. Consequently, even robust upper limits on the total PBH fraction would not invalidate the physical mechanism studied here.

Finally, some of the gravitational-wave events shown in Fig.~\ref{fig:grid} could well be of stellar origin as their mass distributions overlap with those of the PBHs studied in this work. This should be taken into consideration when interpreting the ``best-fit'' values for the spectral index and the asymmetries, although it can be argued that one of those channels will likely dominate. We leave this two-population study, as well as the incorporation of a spectral running, for future work.

\section{Conclusion}\label{sec:conclusion}

If inflation produces a broad fluctuation spectrum with  sufficiently large amplitude at small scales, there is the possibility for an extended PBH mass spectrum to form across the thermal history of the radiation era. Such a spectrum would be imprinted by the cosmic phase transitions occurring in the first minutes following the Big Bang, thus making PBHs a unique probe for the primordial lepton and baryon asymmetry values.
Building on the LAU scenario of Ref.~\cite{Bodeker:2020stj} and on the fully relativistic treatment of PBH formation with a time-dependent equation of state and its peak-theory implementation developed in Refs.~\cite{Escriva:2022bwe,Escriva:2022yaf,Escriva:2023nzn}, we further account for the redistribution of the asymmetries due to neutrino oscillations. We thus follow for the first time the evolution of asymmetries self-consistently from the QCD transition to the neutrino decoupling epoch. This allows us, for different lepton-asymmetry models [see Eqs.~\eqref{LAU:std}--\eqref{LAU:bodeker}], to determine the equation of state of the Universe (Fig.~\ref{fig:EoS}) and compute the PBH spectrum (Fig.~\ref{fig:pbh_spectra}), showing distinct effects of the flavour-dependent asymmetries.
Finally, we determine the GW signal from late PBH merger and find, in agreement with~\cite{Bodeker:2020stj}, that a primordial explanation for observed events (from the GWTC-5 catalogue) prefers a spectral index $n_s\lesssim 0.99$ (see Fig.~\ref{fig:grid}).

Although many observational channels overlap and challenge the extended PBH mass distribution from thermal history accounting for the entirety of dark matter, we motivate this study through the deep connection between PBH mass spectra and fundamental cosmic parameters. If ever detected, the PBH mass spectrum could be one of the keys to decipher the pre-recombination era, with implications from inflation models, to lepto- and baryogenesis and cosmic phase transitions. We argue that constraining methods cannot be applied straightforwardly to an extended PBH mass spectrum and that a PBH population from thermal history is, as things currently stand, too good of a probe to be excluded already. Moreover, the growing numbers of positive evidence in favour of PBHs makes them particularly compelling objects. 

\section*{Acknowledgements}
We thank A.~Kashlinsky for clarifying remarks on the excess source-subtracted near-infrared CIB anisotropies.
M.G.~thanks Oleksii Ivanytskyi and David Blaschke for providing the QCD thermodynamics around the QCD transition. 
M.G.~and G.H.~gratefully acknowledge the financial support provided by the German Federal Ministry of Research, Technology and Space (BMFTR) in the framework of the Knowledge creates perspectives for the region, for the project StStG – DZA – Aufbauphase: Deutsches Zentrum für Astrophysik, Großforschungszentrum in der sächsischen Lausitz: Aufbauphase 2026, grant number 03WSP1746.
J.F.~acknowledges support from the Severo Ochoa Excellence Grant CEX2023-001292-S funded by MICIU/AEI/\allowbreak10.13039/501100011033. A.E.~acknowledges support from the APCTP Junior Group Leader program and JSPS KAKENHI Grant Number 26K17141. A.M.~acknowledges the University of Miami for partial support.

\printcredits

\appendix
\section{Species contribution to the thermodynamics}\label{app:species_contrib}

\begin{figure*}
    \centering
    \includegraphics[width=\linewidth]{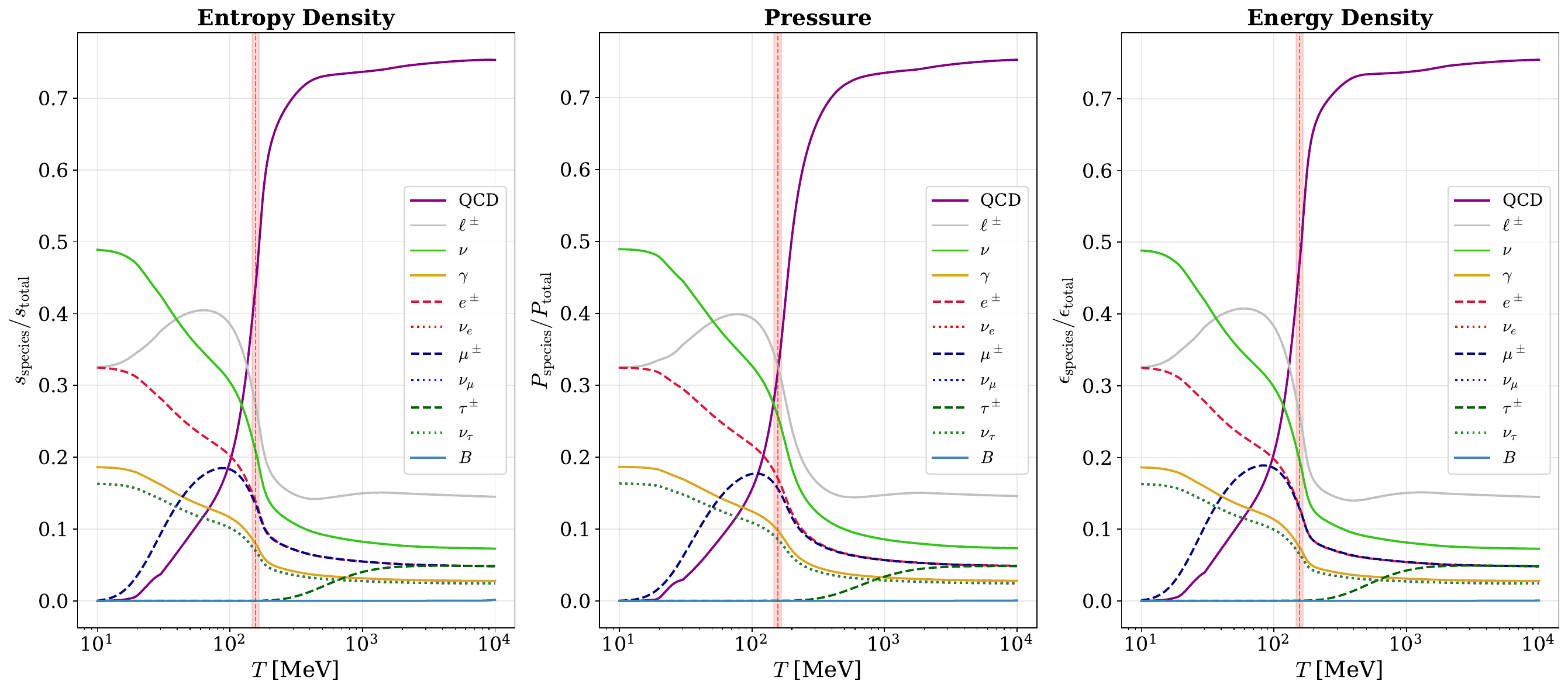}
    \caption{Species-by-species contributions to the entropy density (left), pressure (middle) and energy density (right) between 10 GeV and 10 MeV, for the $\mathrm{(Standard)}$ model. The different species contributions can be read in the legend; ``$B$'' denotes bosons. In this case the neutrino contributions overlap.}
    \label{fig:app:species_contrib_std}
\end{figure*}

\begin{figure*}
    \centering
    \includegraphics[width=\linewidth]{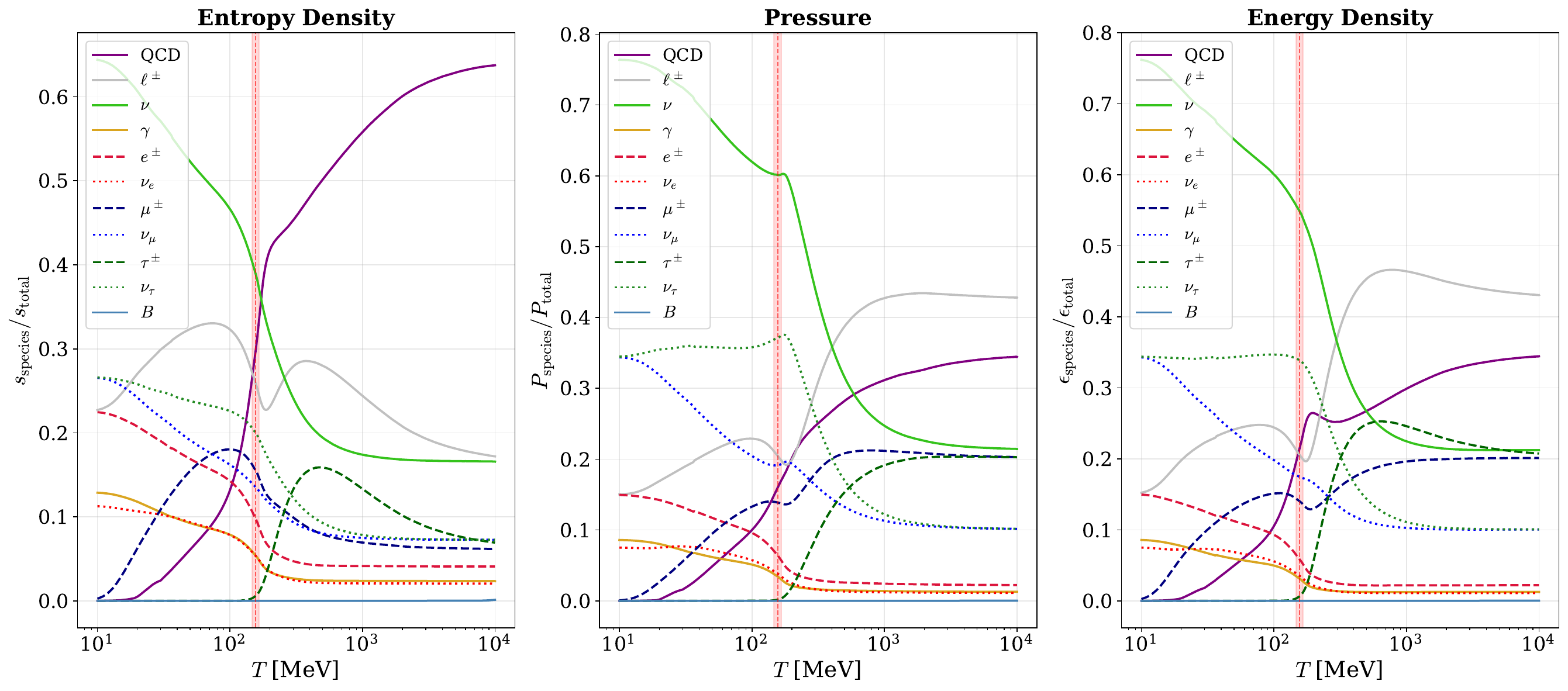}
    \caption{Same as Fig.~\ref{fig:app:species_contrib_std}, for the ${\rm (Type~I-0.1)}$ model.}
    \label{fig:app:species_contrib_type_I_0.1}
\end{figure*}

\begin{figure*}
    \centering
    \includegraphics[width=\linewidth]{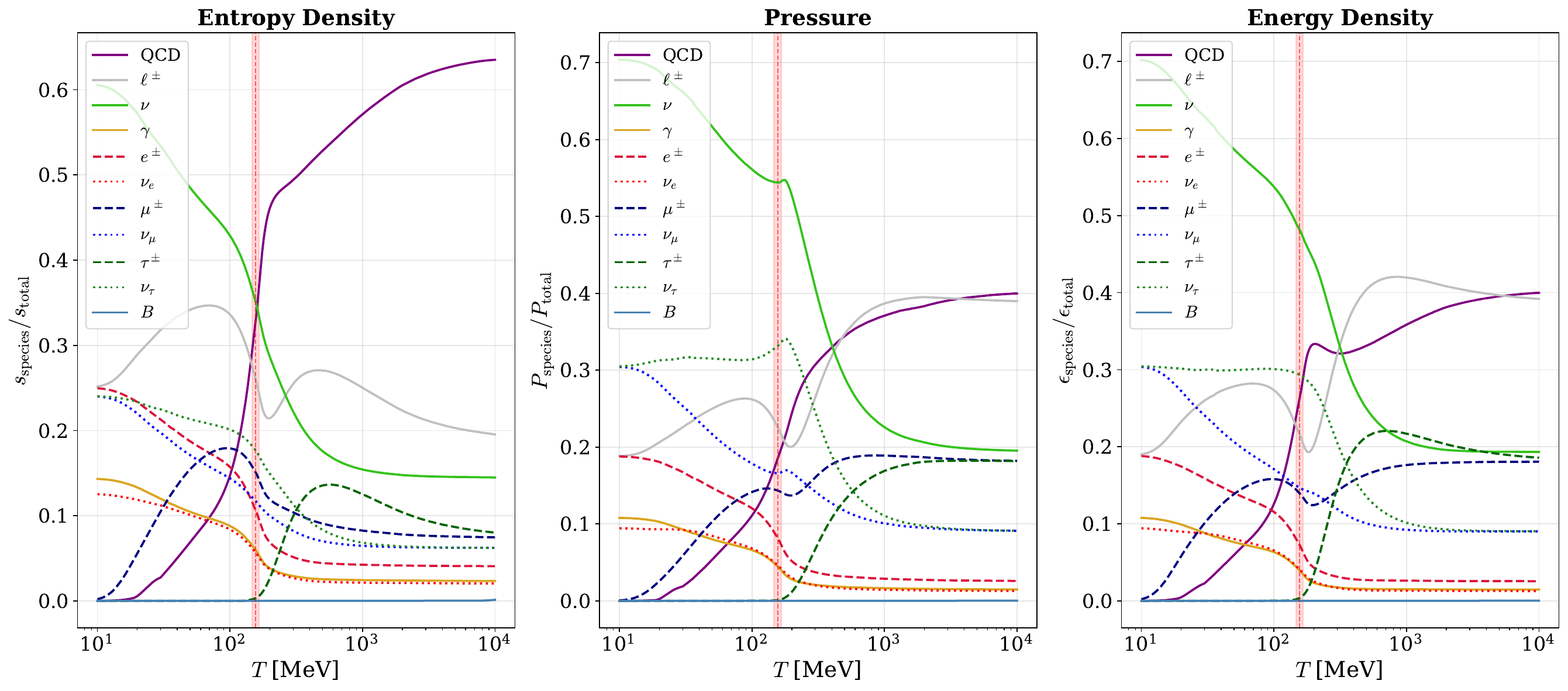}
    \caption{Same as Fig.~\ref{fig:app:species_contrib_std}, for the  ${\rm (Type~I-0.08)}$ model.}
    \label{fig:app:species_contrib_type_I_0.08}
\end{figure*}

\begin{figure*}
    \centering
    \includegraphics[width=\linewidth]{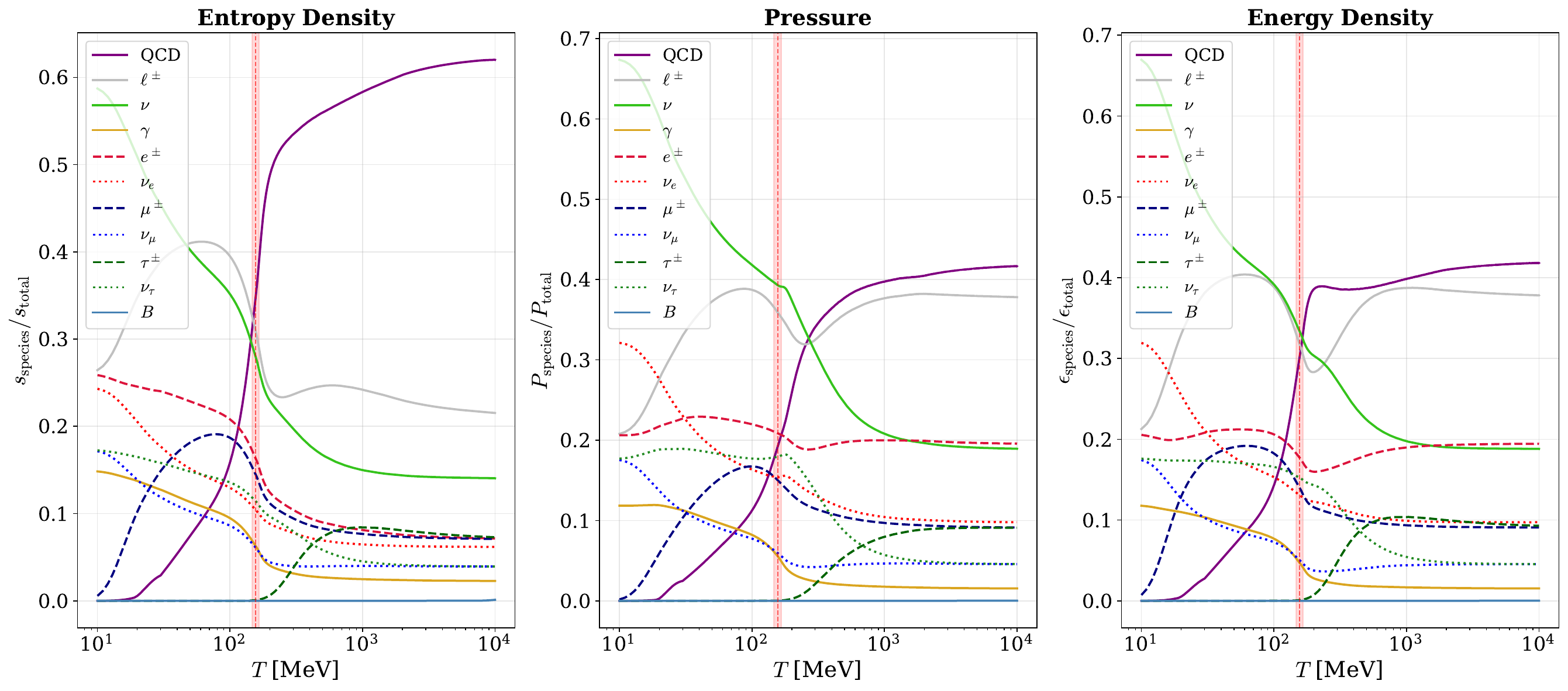}
    \caption{Same as Fig.~\ref{fig:app:species_contrib_std}, for the $\mathrm{(BKOS\text{-}like)}$ model.}
    \label{fig:app:species_contrib_bodeker_like}
\end{figure*}

In this appendix, we show the contributions of the various species to the different thermodynamic quantities (pressure, entropy and energy densities) as computed by 
\texttt{CosmicEoS} and \texttt{Thermal-FIST}, supporting the discussion in Sec.~\ref{subsec:EoS_results}. 
The results for the (Standard), (Type I -- 0.1), (Type I -- 0.08) and (BKOS-like) models are shown in Figs.~\ref{fig:app:species_contrib_std}, \ref{fig:app:species_contrib_type_I_0.1}, \ref{fig:app:species_contrib_type_I_0.08} and \ref{fig:app:species_contrib_bodeker_like}, respectively. 
The connection between \texttt{CosmicEoS} and \texttt{Thermal-FIST} calculations
manifests itself as an inflection point at $T = 37~{\rm MeV}$ in the contribution of the QCD sector, 
reflecting the key difference in the treatment of QCD thermodynamics between the two 
codes. 

It is clear that increasing values of $\ell_\tau$ and $\ell_\mu$ render the QCD sector 
subdominant in pressure and energy density even at $T = 10~{\rm GeV}$. For the entropy 
density, however, the QCD sector remains dominant. This can be understood from the 
relation:
\begin{equation}\label{eq:entropy_density}
    s = (\varepsilon + P - \mu n)/T,
\end{equation}
together with Eq.~\eqref{eq:conservation_equations:b}: with $b = 8.6 \times 10^{-11}$, 
the net baryon number density $n_B$ is extremely small (see Fig.~\ref{fig:asym_evol}), so 
the entropy density of the QCD sector barely changes upon introducing the BAU. Moreover, 
as seen in Fig.~\ref{fig:cosmic_traj}, $|\mu_B|$ is smaller than or comparable to the 
lepton chemical potentials prior to the QCD transition. Thus, with $n_B$ small and $\mu_B$ insufficiently large to 
make the product $\mu_B \times n_B$ significant in Eq.~\eqref{eq:entropy_density}, the 
entropy density of the QCD sector remains essentially unchanged, while that of the leptons 
can be significantly modified by LAU.

At lower temperatures, after the QCD transition, the neutrino 
weight increases with LAU, accounting for up to $\sim 80\%$ of the pressure and energy 
density content. The entropy density, on the other hand, exhibits a much more constant 
behaviour across the different models. The transfer of asymmetry from charged to neutral 
leptons, already seen in Fig.~\ref{fig:asym_evol}, is apparent once again: as soon as the 
contribution of a charged lepton species decreases, the remaining species naturally take a 
larger share of the thermodynamics, but the associated neutrino contribution rises sharply, 
as it must carry the lepton asymmetry.

\section{Neutrino mass ordering and EoS}
\label{app:details_NO_IO}

In this appendix, we provide additional details which support our explanation for the observed differences in the EoS (and, consequently, in the PBH mass spectrum) for temperatures below 10 MeV, as shown in Fig.~\ref{fig:EoS}.

\begin{figure}[!h]
    \centering
    \includegraphics[width=\columnwidth]{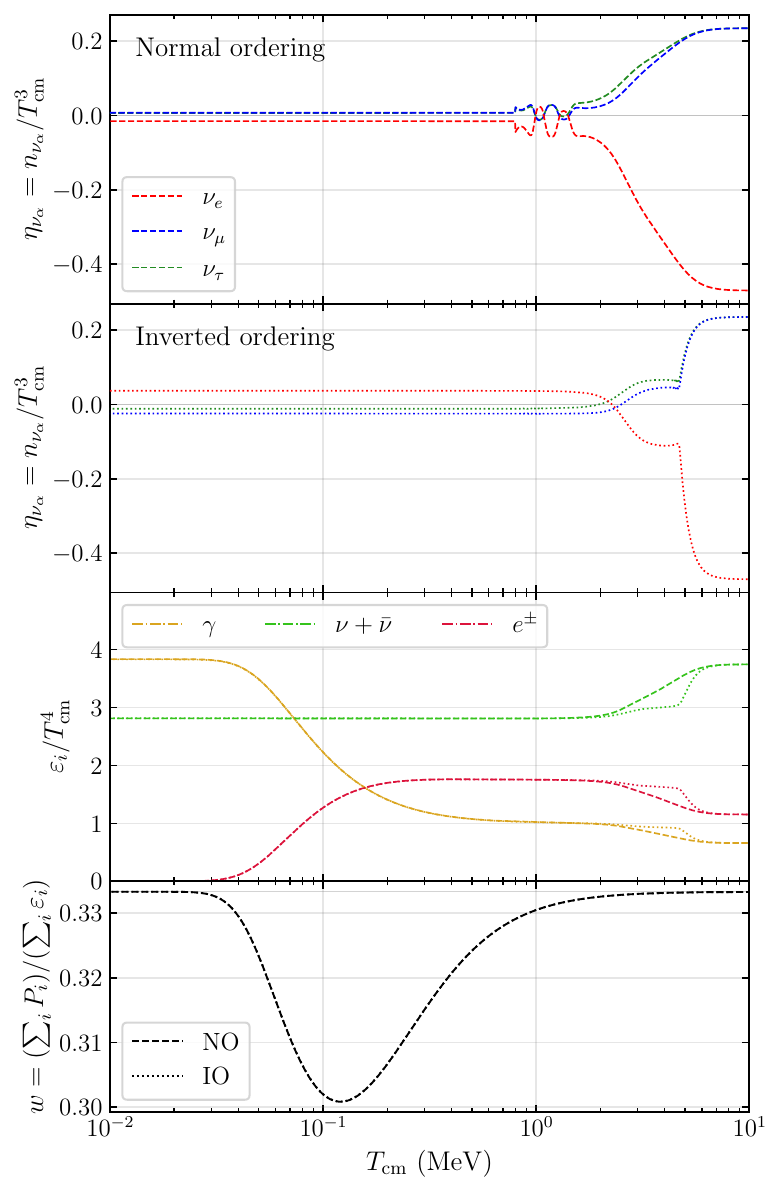}
    \caption{Thermodynamic quantities obtained from the \texttt{NEVO} code, for the $\mathrm{(BKOS\text{-}like)}$ model. From top to bottom: evolution of the asymmetries below $T = 10 \, \mathrm{MeV}$ in the normal ordering of neutrino masses ($\mathrm{NO}$, dashed lines); in the inverted ordering of neutrino masses ($\mathrm{IO}$, dotted lines); individual contributions to the energy density, following the same colour coding as in Fig.~\ref{fig:app:species_contrib_bodeker_like}; and EoS (equivalent to the blue line on Fig.~\ref{fig:EoS}).} 
    \label{fig:app:NEVO_Bodeker}
\end{figure}

\paragraph{(BKOS-like) model.} 
We show in Fig.~\ref{fig:app:NEVO_Bodeker} the thermodynamic quantities obtained from the quantum kinetic equation solver \texttt{NEVO}~\cite{Froustey:2020mcq,Froustey:2021azz,Froustey:2022sla,Froustey:2024mgf} in the (BKOS-like) configuration, between temperatures of 10 MeV and 0.01 MeV.
In the IO case, there is a sharp redistribution of asymmetries at $\sim 5 \, \mathrm{MeV}$, while it is more gradual in the NO case (see top two panels). This results in a net decrease of the energy density of the (anti)neutrino ensemble, and therefore a reheating of the electromagnetic plasma.\footnote{At these temperatures, neutrinos are not yet decoupled and all species are ultrarelativistic, such that $\sum_i \varepsilon_i a^4$ is conserved.} Crucially, although the transient history between [2 MeV, 8 MeV] is different, the different thermodynamic contributions are indistinguishable below 1 MeV. This is why the EoS shows no visible difference (see bottom panel).

\begin{figure}[!h]
    \centering
    \includegraphics[width=\columnwidth]{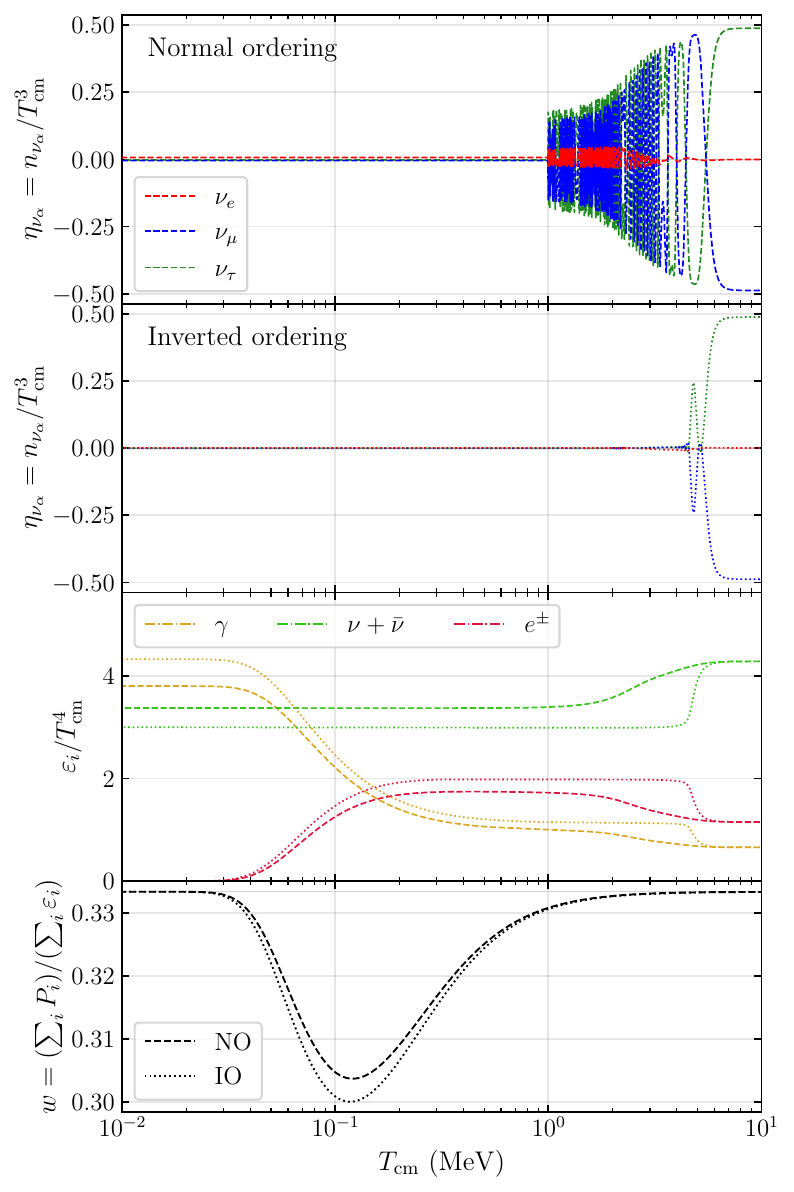}
    \caption{Same as Fig.~\ref{fig:app:NEVO_Bodeker}, for the ${\rm (Type~I-0.08)}$ model. The results are similar for the ${\rm (Type~I-0.1)}$ model.}
    \label{fig:app:NEVO_TypeI_0.08}
\end{figure}

\paragraph{(Type I -- 0.08) model.} The story is quite different in the (Type I) cases, and we focus in Fig.~\ref{fig:app:NEVO_TypeI_0.08} on the (Type I -- 0.08) model. While in the NO case the asymmetries are gradually reduced, they are completely and rapidly washed out in the IO case. There is therefore a larger redistribution of $\varepsilon_i$ in the IO case. The transfer is also less efficient in the NO case since it takes place at lower temperatures, hence when neutrinos have started decoupling. We note that, even though there seems to be a sharp numerical transition at $T_\mathrm{cm} = 1 \, \mathrm{MeV}$ in the top panel, which is associated with the \texttt{NEVO} solver switching to an adiabatic method neglecting the self-interaction potential~\cite{Froustey:2021azz,Froustey:2024mgf}, the absence of sharp features in the energy density panel ensures that the thermodynamics relevant for the EoS are well captured.

As a consequence of this different asymmetry equilibration, and as shown in the third panel, the net contribution of $\nu/\bar{\nu}$ to the total energy density is visibly larger in the NO case. This drives the EoS towards $w = 1/3$, consistent with the bottom panel (see also Fig.~\ref{fig:EoS}). Finally, we note that the higher temperature of the electromagnetic plasma for a given $T_\mathrm{cm}$ explains the small shift of the dip in the bottom panel, a feature already explained when comparing with the (Standard) case in Sec.~\ref{subsec:EoS_results}.

\section{Modification of the spectral index}\label{app:spectral_index}

\begin{figure*}
    \centering
    \includegraphics[width=\linewidth]{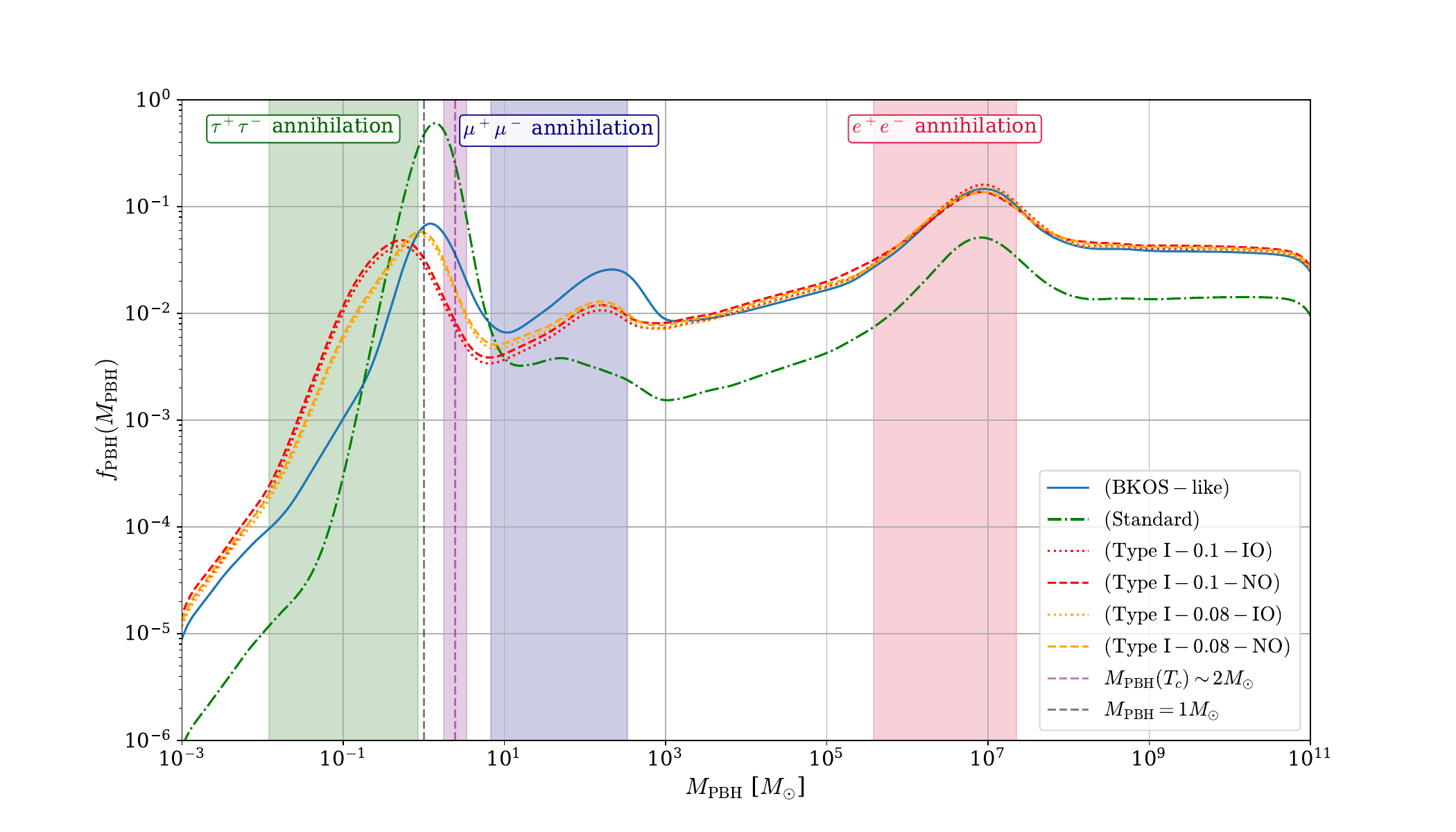}
    \caption{Same as Fig.~\ref{fig:pbh_spectra}, but with a spectral index $n_s=0.94$.}
    \label{fig:app:pbh_spectra_ns_0.94}
\end{figure*}

\begin{figure*}
    \centering
    \includegraphics[width=\linewidth]{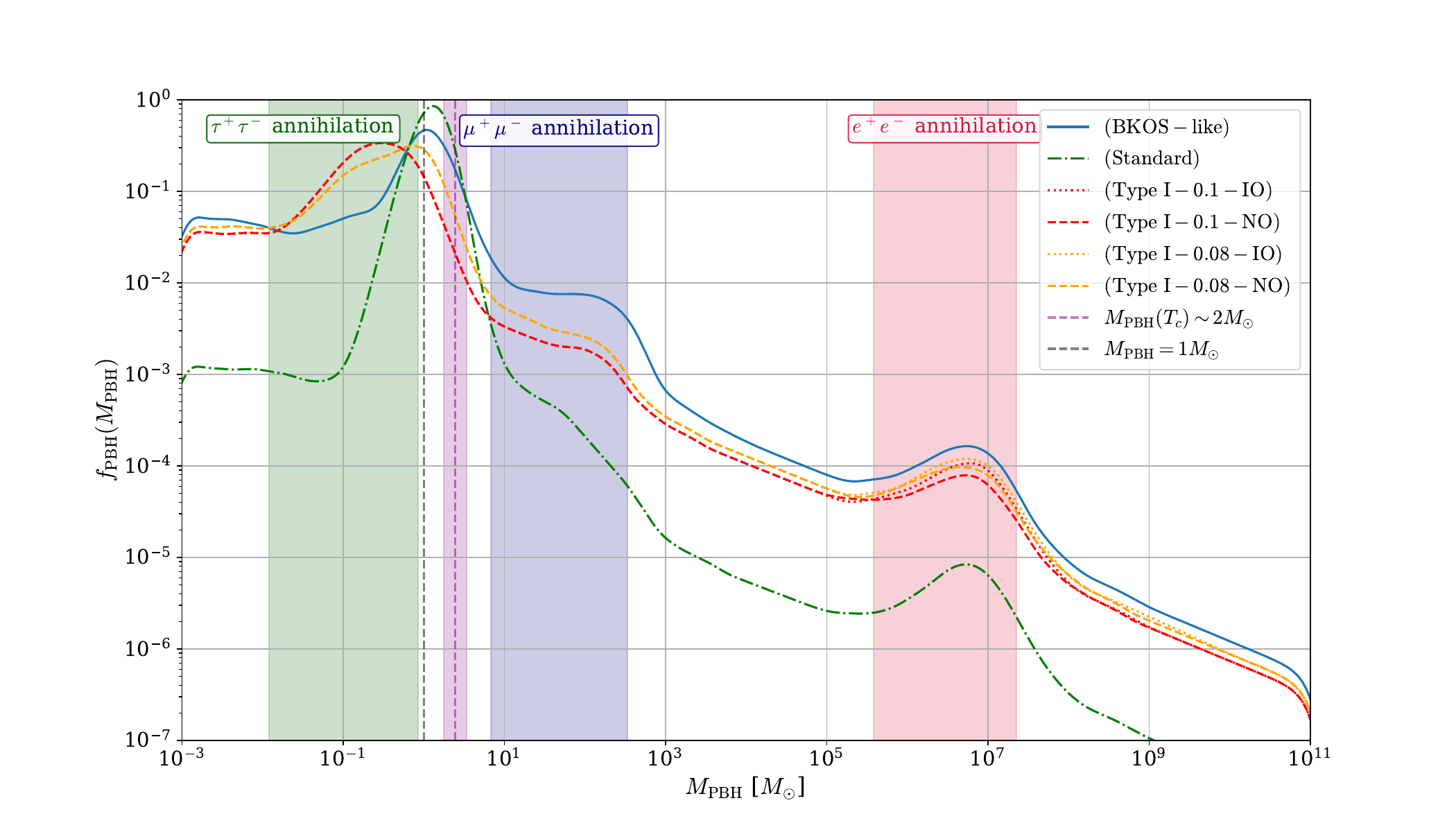}
    \caption{Same as Fig.~\ref{fig:pbh_spectra}, but with a spectral index $n_s=0.99$.}
    \label{fig:app:pbh_spectra_ns_0.99}
\end{figure*}

\begin{figure*}
    \centering
    \includegraphics[width=\linewidth]{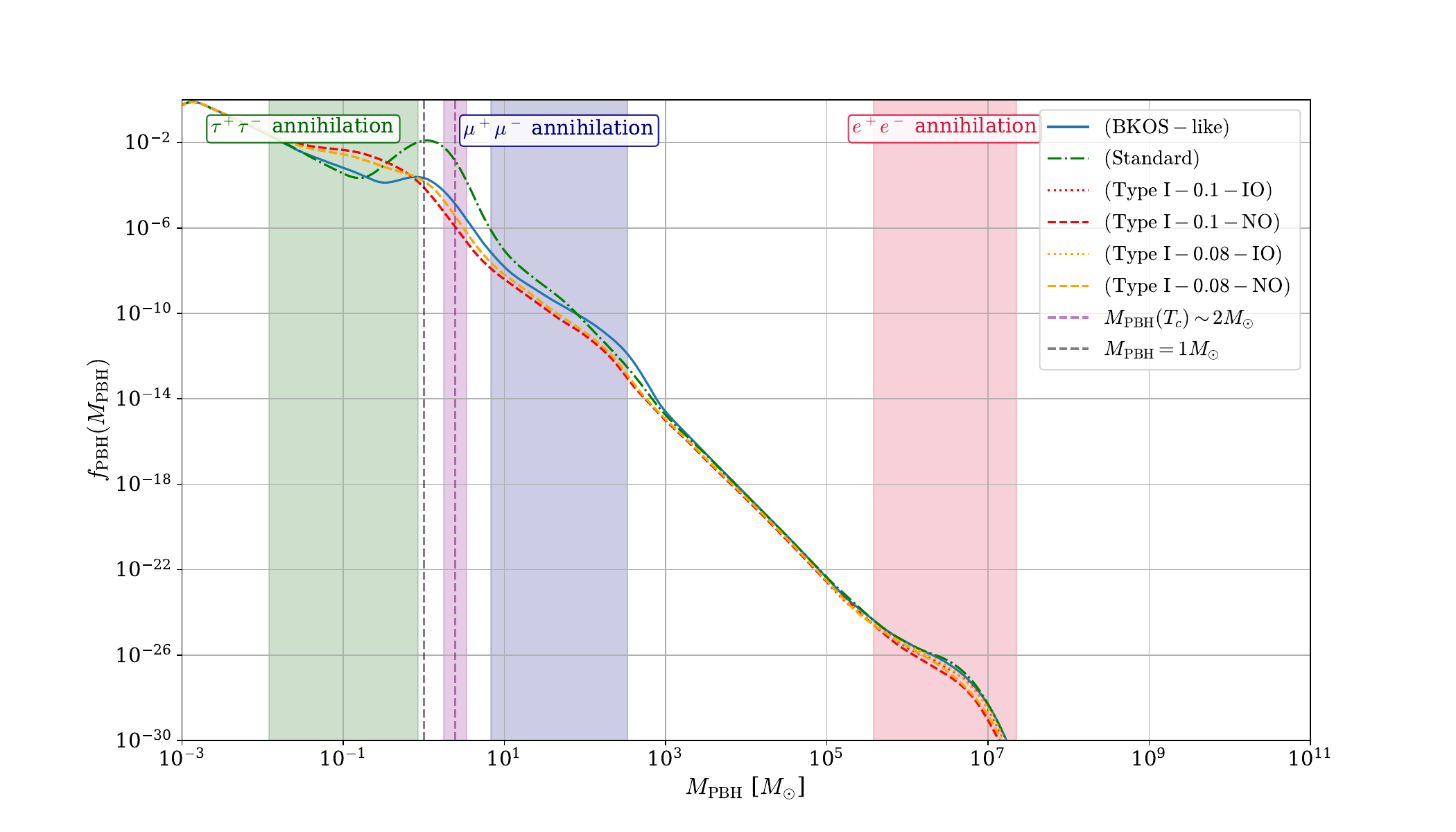}
    \caption{Same as Fig.~\ref{fig:pbh_spectra}, but with a spectral index $n_s=1.1$.}
    \label{fig:app:pbh_spectra_ns_1.1}
\end{figure*}

In this appendix we plot the PBH mass spectra for different values of $n_s$. Increasing 
the value of $n_s$ tilts the curves and makes the heavy PBH contribution less significant, 
while decreasing $n_s$ can make the $e^+e^-$ annihilation peak the dominant one for the 
lepton-asymmetric models; see Figs.~\ref{fig:app:pbh_spectra_ns_0.94}--\ref{fig:app:pbh_spectra_ns_1.1}. 

For $n_s = 0.94$, the ${\rm (Type~I\text{-}0.08)}$ 
model shows an a priori peculiar behaviour, as differences appear between the NO and IO cases in the low-mass range of the spectrum prior to the $e^+e^-$ annihilation peak. This could be surprising, since the EoS only differs between the NO and IO cases in the $e^+ e^-$ annihilation region, see Fig.~\ref{fig:EoS}. The reason lies in the common normalisation to 
$f_{\rm PBH}^{\rm tot} = 1$: for $n_s = 0.94$ the dominant peak shifts to the $e^+e^-$ 
annihilation peak, and when using the same normalisation, a slight mitigation of the 
dominant peak forces the rest of the distribution to contribute more.

In Fig.~\ref{fig:app:pbh_spectra_ns_1.1} with $n_s = 1.1$ the spectra shown are extremely tilted and the structure features associated with the thermal history do not dominate the distributions anymore. 

\clearpage

\bibliographystyle{elsarticle-num}

\bibliography{refs}

\end{document}